\documentclass[twocolumn]{aastex631}
\usepackage{blindtext}
\usepackage{amsmath}
\usepackage{bm}
\usepackage{graphicx}
\usepackage{svg}
\shorttitle{}
\shortauthors{}

\begin{document}

\title{Planetesimal-Driven Instabilities in Resonant Chains of Cold Neptunes and Their Dynamical Outcomes}

\author[0009-0001-1504-9232]{Ryan LoRusso} 
\affiliation{Department of Astronomy, Indiana University, Bloomington, IN 47405, USA}
%\email{rlorusso@iu.edu}

\author[0000-0003-0412-9314]{Cristobal Petrovich}
\affiliation{Department of Astronomy, Indiana University, Bloomington, IN 47405, USA}
%\email{cpetrovi@iu.edu}

\author[0000-0002-5181-0463]{Hareesh Bhaskar} 
\affiliation{Department of Astronomy, Indiana University, Bloomington, IN 47405, USA}
%\email{bhareeshg@gmail.com}

\begin{abstract}

Cold Neptunes and sub-Neptunes are among the most common products of planet formation and likely dominate the angular-momentum budgets in most planetary systems; yet their dynamical impact on planetary architectures remains poorly understood.
Using N-body simulations, we investigate the evolution of multi-Neptune systems assembled into resonant chains during the gas-disk phase and later coupled to remnant planetesimal disks.
We show that planetesimal disks containing $\simeq 1$--$4\%$ of the planetary mass efficiently disrupt resonant chains and trigger global dynamical instabilities on timescales of $1~\mathrm{Myr}$--$1~\mathrm{Gyr}$, providing a pathway for delayed instability long after gas-disk dispersal, albeit with instability timescales that are highly sensitive to disk mass.
The ensuing instability drives large-scale orbital rearrangement and loss of planets through collisions, tidal disruption, and ejections. 
Notably, in most systems at least one planet is scattered inward to $\sim 0.1~\mathrm{au}$ on $\sim 10$--$100$ Myr timescales (for $\sim 5$--$50\; M_\oplus$ planets) following instability onset, with a substantial fraction undergoing tidal capture or disruption. 
This tidal capture can provide a natural pathway to hot Neptune formation, while compact inner chains, if present, would be destroyed on $\sim 100~\mathrm{Myr}$ timescales by cold sub-Neptunes, naturally explaining the observed decline in the resonant fraction.
We argue that the predictions of our model, which yields mass-segregated planets and corresponding relative abundances of cold, wide-orbit, and free-floating planets, can be tested by ongoing and upcoming microlensing surveys.
\end{abstract}

\section{Introduction} \label{sec:Introduction}
\subsection{The Dominance of Cold Neptunes}

 {\it Kepler} revealed that planets with Neptune-like radii are far more abundant than Jupiter-size planets on orbits interior to $\sim 1\;\mathrm{AU}$ \citep{Hsu2019}. 
 Combined with radial-velocity measurements, these results imply that Neptune-like planets—here defined as planets with masses of $\sim 10$–$30 M_{\oplus}$—are significantly more common than Jupiter-mass planets at these separations \citep{Wright2012}. 
 Radial-velocity surveys further suggest that this dominance likely extends to longer orbital periods \citep{Mignon2025}, although current constraints remain limited by sensitivity and observational biases.

At wider separations ($a \sim$ a few~au), gravitational microlensing provides the strongest constraints on Neptune-mass planets. 
Microlensing surveys consistently find that cold Neptunes are significantly more common than cold Jupiters \citep{Zang2025}, reinforcing the conclusion that Neptune-mass planets dominate the intermediate-to-wide orbital regime. 
These results imply that while only $\sim 10\%$ of Sun-like stars host cold Jupiters \citep{Fulton2021}, the majority of planetary systems may host cold Neptunes as their most massive planets. As a consequence, Neptune-mass planets may often control the angular-momentum budget of planetary systems and play a dominant role in shaping their long-term dynamical evolution.

\subsection{Observational Evidence for Dynamical Instabilities of Neptunes}
The large inferred population of free-floating Neptune-mass planets \citep{Sumi2023} suggests that dynamical scattering and ejection processes operate efficiently in planetary systems hosting Neptunes. While ejections are more readily produced by massive Jovian planets, the low occurrence rate of cold Jupiters implies that they cannot be the primary source of the observed free-floating population \citep{Fulton2021,HaddenWu2025}. Accounting for the inferred abundance of free-floating Neptunes using only Jupiter-mass perturbers would require an implausibly large number of ejections—on the order of $\gtrsim 10$ Neptune-mass planets per Jupiter-hosting system \citep{HaddenWu2025}. Given the high intrinsic occurrence rate of Neptune-mass planets, it is therefore natural to expect that Neptunes themselves frequently eject one another through mutual dynamical instabilities. 

Additional evidence for violent dynamical histories emerges from the short-period regime. Analogous to hot Jupiters, an emerging population of hot Neptunes with orbital periods of $\sim 3$–$6$ days has been identified \citep{CastroGonzalez2024}. These planets frequently exhibit large stellar obliquities \citep{Dai2017} and, in some cases, residual eccentricities \citep{Hixenbaugh2023,Dugan2025}, pointing to a history of strong dynamical excitation followed by high-eccentricity tidal migration \citep{Dawson_HT_review,Bourrier2023_hot_Neptunes}. 

The excitation of extreme eccentricities is typically attributed to massive outer companions---cold Jupiters or stellar companions. However, long-term radial-velocity monitoring rules out the presence of cold Jupiters in most hot Neptune systems \citep{Espinoza-Retamal2024,Handley2026}. In contrast, Neptune-mass companions at wider separations often fall below current detection limits. Given the high intrinsic occurrence rate of Neptune-mass planets and the relative scarcity of cold Jupiters, Neptune–Neptune interactions provide a natural pathway for driving the dynamical excitation required for high-eccentricity migration.

Taken together, these observations indicate that dynamical instabilities among Neptune-mass planets are likely common and may play a central role in shaping both bound and unbound Neptune populations. What physical processes initiate these instabilities, and on what timescales do they emerge following gas-disk dispersal?

\subsection{Planetesimal-Driven Migration and Delayed Instability}

During the gas-disk phase, interactions with the protoplanetary disk naturally drive orbital migration and eccentricity damping, frequently leading to capture into mean-motion resonances and the assembly of compact resonant chains \citep{Izidoro2017,Goldberg2025,Ogihara2026}. Transit surveys support this picture, revealing a significant excess of multi-planet systems near orbital commensurabilities, as well as widespread transit-timing variations in young systems with ages of $\sim 10$--$100~\mathrm{Myr}$ \citep{Dai2024,Lopez2026_TTV}. 

In contrast, among older systems the fraction of planets residing in resonant chains drops sharply, with fewer than $\sim 10\%$ of systems remaining resonant. This decline indicates that most resonance chains are disrupted on timescales long after the gas disk has dispersed, yet the physical mechanism responsible for this delayed resonance breaking remains uncertain \citep{Li2025}.

A natural candidate for driving this evolution is a remnant planetesimal disk. Planetesimal populations are expected to be common, as suggested by the high incidence of debris disks, particularly around young stars \citep{Moor2016_debris,Montesinos2016_debris}. Gravitational scattering and ejection of planetesimals exchange angular momentum with planets, producing planetesimal-driven migration. In two-planet systems, this process is known to generate divergent migration and resonance breaking \citep{Chatterjee2015,Ghosh2023_repulsion,Wu_pingpong2024}. How planetesimal-driven migration operates in higher-multiplicity systems of cold Neptunes---beginning from disk-assembled resonant chains---remains largely unexplored\footnote{Only recently, \citet{HaddenWu2026} and \citet{Choksi2026} explored resonance chain breaking by planetesimal disks, focused on short-period planets.}. In such systems, collective interactions can amplify small perturbations, potentially triggering large-scale dynamical instabilities \citep{Pichierri2020_resonance,Izidoro2021_breaking,Goldberg2022_stab_res}.

\subsection{Consequences of Instabilities}

The dynamical outcomes of planetary instabilities depend sensitively on the Safronov number\footnote{Here the parabolic Safronov number is used, where the relative velocity between bodies is approximately the Keplerian $v_{orb}$. See also \citet{Tremaine2023} pages 442 \& 454.},
\begin{equation}
\label{eqn:safronov}
\Theta = \frac{v_{\rm esc}^2}{2 v_{\rm orb}^2},
\end{equation}
which measures the relative importance of gravitational scattering compared to physical collisions. 

In systems dominated by cold Jupiters ($\Theta \gg 1$), close encounters typically lead to strong gravitational scattering and ejections rather than mergers. Dynamical instabilities in such systems have been widely studied and successfully reproduce the broad eccentricity distribution of giant exoplanets through planet–planet scattering followed by ejections \citep{Chatterjee,JT2008}. 

At the opposite extreme, compact systems of inner super-Earths and sub-Neptunes are expected to reside in a regime with $\Theta \ll 1$, where physical collisions dominate over ejections. Instabilities in this limit tend to produce mergers that reduce multiplicity while leaving behind dynamically packed systems close to the stability boundary \citep{PuWu2015,Obertas2017,Goldberg2022_stab_res}. 

Cold Neptune systems occupy an intermediate regime with $\Theta \sim 1$, where mergers, scattering, and ejections can occur with comparable probability. Despite the observational evidence suggesting that Neptune-mass planets are common and frequently unstable, this regime has received comparatively little attention. Recent studies have examined Neptune–Neptune interactions in the context of producing free-floating planets \citep[e.g.,][]{HaddenWu2025}, but often neglect important physical processes such as planet–planet collisions or tidal disruptions during close stellar passages.

As a result, the dynamical consequences of instabilities in Neptune-dominated systems remain poorly understood. In particular, it is unclear how such instabilities redistribute planets between inner, outer, and unbound populations, or how frequently they produce outcomes such as tidal capture, tidal disruption, or wide-orbit survivors.

\medskip

In this work, we postulate that planetary embryos may be able to drive instabilities in resonant chains of cold Neptunes, and investigate the dynamical outcomes of destabilized systems.
We structure the paper as follows. In \S\ref{sec:setup}, we describe the initial conditions and methods of our N-body cold Neptune simulations. In \S\ref{sec:results}, we provide the results of these simulations. In \S\ref{sec:discussion}, we discuss the implications of our proposed model and identify areas in which further study is required. In \S\ref{sec:conclusion} we end by reviewing the key ideas presented in this work.

\section{Scattering experiments} \label{sec:setup}

We use the N-body integration code REBOUND to perform all simulations discussed in this work \citep{REBOUND}. We use REBOUNDx code \citep{REBOUNDx} to add physical damping prescriptions for modeling resonant capture (via \texttt{modify\_orbits\_direct}) with the full $p=1$ coupling of \citet{Deck2015}. Importantly, we do not include a prescription for tidal dissipation in these scattering experiments. For our numerical integrations we use a Gragg-Bulirsch-Stoer integrator (as implemented in REBOUND, which adapts the method of~\citet{BS_algorithm}) with a maximum timestep of 0.1 years, and adopt the default absolute and relative error thresholds of $10^{-8}$.   

We track losses of simulated planets and planetesimals across three outcomes: \textit{collision} with another planet ($C$), \textit{ejection} from the system ($E$), and tidal \textit{disruption} or collision with the star central to the system ($D$). 
We detect collisions via the `line' method in REBOUND (in which celestial bodies collide if following a straight-line path between their positions at the start and end of a timestep collision would occur). 
Collisions between planets and planetesimals we resolve via a custom resolution function; mass and momentum are conserved and the radius is determined using the mass-radius relation of \citet{Muller2024} with a small upwards correction\footnote{The mass-radius relation is scaled such that a Neptune-massed planet has the radius of Neptune, corresponding to a uniform increase of 3.2\% in radius.}.
When a body exceeds 1000 AU in distance from the center of mass of the system, it is considered ejected.
When a body crosses within the Roche limit of the star, we treat it as a tidal disruption\footnote{This condition is achieved by increasing the radius of the star to the Roche limit and treating the tidal disruptions as stellar collisions.  
See also \S\ref{subsec:cn_disrupt}.}. 
In principle, these so-called `disruptions' may be tidal captures since we include no prescription for tidal forcing.

We consider a planetary system to be actively evolving if the semimajor axes of any planets change by more than 1\% over the last 100 Myr of its evolution or if perihelion drops below 0.1 AU during those 100 Myr.
We also examine two-planet systems using well-tested stability criteria from the literature. 

The structure of our simulations consists of five broad phases.
In the first phase, following the procedure of \citet{Tamayo2017}, we induce resonant capture of a series of planets by prescribing disk migration for only the outermost planet in the series, leading to the formation of a resonant chain. 
All planets are given eccentricity damping, but only the outermost is given semi-major axis damping. 
We select damping timescales $\tau_a,\tau_e$ to both achieve a resonant capture of all five planets into 3:2 resonance within the timescale of the disk lifetime and to ensure that the planets be captured deep into such a resonance. 
We prescribe $\tau_a=4~\rm{Myr}$, equal to the duration of the first phase, to produce slow migration. 
Except where otherwise mentioned, we prescribe eccentricity damping for the disk such that the parameter $K = \tau_a /\tau_e = 100$\footnote{The $K$ parameter appears as a constant ratio relating the strength of the eccentricity damping to the strength of the semimajor axis damping in~\citet{Lee2002}; see also \S2 of \citet{Tamayo2017} for remarks on the effects of selecting such $K$.}. These timescales are expected to produce 3:2 resonance capture (see e.g., the criteria summarized in \citet{Batygin2026}). 

The chosen values of $\tau_a,\tau_e$ for Neptune-massed planets are equivalent to a disk with aspect ratio $({h}/{r}) \sim 0.04$ and surface density $\Sigma \sim 32 \mathrm{~g~cm^{-2}~} \left( {r}/{1~\mathrm{au}} \right)^{-3/2}$ using the equations of \citet{Tanaka2002} and \cite{Tanaka2004} for chosen slope of the density profile $\alpha = -3/2$. This is dramatically lower than the $\sim 1700 \mathrm{~g~cm^{-2}~} \left( {r}/{1~\mathrm{au}} \right)^{-3/2}$ of the classic solar disk model of \citet{Hayashi1981}\footnote{Although this seems to imply that a very sparse disk must be required to achieve the desired resonance capture, we assume damping applies only to one planet. If we account for the migration of both planets and treat the prescribed $\tau_a$ as arising from the differences in their migration rates, we instead find a disk surface density of $\Sigma \sim 400 \mathrm{~g~cm^{-2}~} \left( {r}/{1~\mathrm{au}} \right)^{-3/2}$.}.

In the second phase, we approximate disk dispersal by exponentially increasing the damping timescales (and thus exponentially reducing the strength of the damping) for three e-folding times over 1 Myr. 

In the third phase, we remove damping and introduce planetesimals in-situ into the system. The system is then evolved for 2 Myr. 
We let simulations continue for 100 Myr (fourth phase) and then a further 1 Gyr (fifth phase). Simulation snapshots are saved every 2 kyr in the third phase and every 100 kyr in the fourth and fifth phases.
While there are no other changes to simulation parameters between the third, fourth, and fifth phases, the distinction is useful as it permits focus on the collapse of resonant angles in the third phase (in practice, simulated systems fall out of resonance very quickly, on timescales $\ll2$ Myr), typical destabilization of systems in the fourth phase, and the long-term evolution towards a stable state in the fifth. A typical simulation requires $\sim 20$ to over $120$ CPU hours, depending on the suite.

\subsection{Initial Conditions} 
\label{subsec:ic}

Our archetypal series of simulations is our \textit{f5} simulations\footnote{In our naming convention, the letter at the beginning indicates the outermost planet index (i.e., in the archetypal series of simulations, \textit{f} indicates a five-planet system) and the number immediately following it indicates the sum of the integers of the first or second-order resonance in question (here \textit{5}, or a 3:2 resonance). Such a sum will be unique for any first or second-order resonance.}. 
For our \textit{f5} series of simulations, we assume initial conditions of a solar mass star and five identical, zero eccentricity, Neptune-mass planets. The innermost of the cold Neptunes were started at semimajor axis $a_1 = 4~\mathrm{AU}$. All planets in the chain were started just wide of the 3:2 ratio of commensurability; measured for our \textit{f5} series of simulations, using a fractional ``resonance offset" parameter $\Delta$ (for the $p:q$ MMR) of
\begin{equation}
\label{eqn:resonance_offset}
\Delta = \left (\frac{q P_{\rm outer}}{p P_{\rm inner}} \right) - 1,
\end{equation}
planets were started with $\Delta = 0.033$; Equation~\ref{eqn:resonance_offset} is equivalent to formulations of $\Delta$ provided in prior works~\citep{Lee2013,Silburt2015,Ramos2017,dai2023}. This is sufficient for all Neptunes to be instantiated outside of resonance.

The planets are assumed to have inclinations uniformly randomly distributed between 0 and 1 degrees, with longitude of the ascending node and the longitude of pericenter uniformly randomly distributed across their entire range.

Our planetesimals are assumed to have uniform masses and represent the upper limits to the mass spectrum of planetesimals in the disk---given their masses, they might better be described as planetary embryos. Due to their size, we model our planetesimals as full particles. We vary the characteristics of planetesimals between \textit{f5} simulations; the series of the simulations is followed by a label of the form \textit{.AnBm}. 
\textit{A} represents the number of planetesimals per planet in the system, and \textit{B} the total mass of the planetesimals as a percentage of the total mass of planets in the system. 

Our fiducial simulation set is the \textit{f5.6n2m} simulations, which corresponds to thirty planetesimals totaling 2\% of the mass of the five Neptune-mass planets of the system. For reference, these are very nearly half Mars-massed planetesimals. 
For all simulations, the semi-major axes of the planetesimals are drawn from a log-uniform distribution whose bounds are set such that the innermost Neptune has a semimajor axis \textit{s} times that of the innermost possible planetesimal, and the outermost planetesimal has a semimajor axis \textit{s} times that of the outermost Neptune. For all simulations in this work we use $s = 1.2$.
The initial eccentricities for the planetesimals are uniformly drawn between 0 and 0.05. Otherwise their orbital parameters are drawn as the planets'. 

In order to consider the impact of initial planetary multiplicity on results, we simulate systems containing three and four identical zero-eccentricity Neptune-mass planets with the initial conditions otherwise described. These are our \textit{d5} and \textit{e5} series of simulations respectively. We remark that because we initialize planetesimals and their individual masses as proportional to the number of planets in the system, both the individual planets and planetesimals in our \textit{d5.6n2m} and \textit{e5.6n2m} are of equal mass to those in the \textit{f5.6n2m} simulations. However, the outermost extent of the initialized planetesimal disk is truncated as the outermost planet is not as distant as in the \textit{f5} series.

We also simulate two suites of five-planet 3:2 systems to determine the dependence of our results on planetary mass. In these suites the masses of the planets are increased or decreased by a factor of three. To distinguish these suites from the \textit{f5} series, they are preceded by a prefix \textit{mm(+3)} or \textit{mm(-3)} respectively, indicating a modified mass and the factor by which that mass is modified. In these simulations to maintain 3:2 resonance capture we prescribe $K = 200$.

\begin{table*}[!bth]

\begin{center}
\caption{Summary of simulated systems and planetary outcomes}
\label{table:summary_final}

\begin{tabular}{r|ccccc|c||c|ccccc|ccc}

\multicolumn{16}{l}{\textsc{Three-planet simulations}}\\
\hline
\hline
Suite&$N_{pl}$&$\overline{\log_{10}{\tau_{dest.}}}$&$a_1$ [AU]&$K$&$\epsilon$&$N_{sys}$&$N_{\rm active}$&5 p.&4 p.&3 p.&2 p.&1 p. & C & D & E \\

\hline
\multicolumn{16}{l}{\textsc{1.102 Gyr, 1 $M_{Nep}$}}\\
\hline

{\it d5.6n2m}    & 18  & $8.28 \pm 0.76$ & 4 & 100 & 0.02  & 30 & 2
    & - & - & 17 & 9 & 4 & 9 & 4 & 4 \\

\hline

\multicolumn{16}{l}{}\\
\multicolumn{16}{l}{\textsc{Four-planet simulations}}\\
\hline
\hline
Suite&$N_{pl}$&$\overline{\log_{10}{\tau_{dest.}}}$&$a_1$ [AU]&$K$&$\epsilon$&$N_{sys}$&$N_{\rm active}$&5 p.&4 p.&3 p.&2 p.&1 p. & C & D & E \\

\hline
\multicolumn{16}{l}{\textsc{1.102 Gyr, 1 $M_{Nep}$}}\\
\hline

{\it e5.6n2m}    & 24 & $7.48 \pm 0.65$    & 4 & 100 & 0.02  & 30 & 5
    & - & 2 & 2 & 21 & 5 & 26 & 13 & 20 \\

\hline

\multicolumn{16}{l}{}\\
\multicolumn{16}{l}{\textsc{Five-planet simulations}}\\
\hline
\hline
Suite&$N_{pl}$&$\overline{\log_{10}{\tau_{dest.}}}$&$a_1$ [AU]&$K$&$\epsilon$&$N_{sys}$&$N_{\rm active}$&5 p.&4 p.&3 p.&2 p.&1 p. & C & D & E \\

\hline
\multicolumn{16}{l}{\textsc{1.102 Gyr, 0.333 $M_{Nep}$}}\\
\hline

{\it mm(-3)f5.3n4m}    &15  & $7.28 \pm 0.47$   &4 & 200 &0.04 & 30 & 18
    & 0 & 1 & 15 & 14 & 0 & 32 & 6 & 35 \\

\hline
\multicolumn{16}{l}{\textsc{1.102 Gyr, 1 $M_{Nep}$}}\\
\hline

{\it f5.0n0m}    &0    &$>t_{max}$    &4 & 100 &N/a  & 15 & 0
    & 15 & 0 & 0 & 0 & 0 & 0 & 0 & 0 \\

\hline

{\it f5.3n1m}    &15  & $7.70 \pm 0.36$ &4 & 100 &0.01  &30 & 12
    & 0 & 0 & 3 & 24 & 3 & 12 & 50 & 28 \\
{\it f5.3n2m}    &15  & $6.63 \pm 0.38$ &4 & 100 &0.02  &30 & 9
    & 0 & 0 & 9 & 19 & 2 & 22 & 34 & 27 \\
{\it f5.3n4m}    &15  & $6.06 \pm 0.41$   &4 & 100 &0.04 &30 & 14
    & 0 & 0 & 4 & 22 & 4 & 26 & 35 & 29 \\
{\it f5.6n1m}    &30  & $8.42 \pm 0.33$   & 4 & 100 &0.01  &30 & 13
    & 8 & 0 & 6 & 13 & 3 & 15 & 20 & 28 \\
{\it f5.6n2m}    &30  & $7.27 \pm 0.37$   & 4 & 100 &0.02  &30 & 11
    & 0 & 0 & 4 & 24 & 2 & 21 & 35 & 32 \\
{\it f5.6n4m}    &30  & $6.26 \pm 0.45$   &4  & 100 &0.04  &30 & 10 
    & 0 & 0 & 6 & 19 & 5 & 20 & 40 & 29 \\
{\it f5.12n1m}    &60  & $8.90 \pm 0.18$  & 4  & 100 &0.01  &30 & 7 
    & 25 & 2 & 0 & 3 & 0 & 0 & 1 & 10 \\
{\it f5.12n2m}    &60  & $7.86 \pm 0.48$   &4  & 100 &0.02  &30 & 14
    & 0 & 1 & 8 & 16 & 5 & 15 & 24 & 46 \\
{\it f5.12n4m}    &60  & $6.96 \pm 0.47$   &4 & 100  &0.04  &30 & 9
    & 0 & 0 & 5 & 25 & 0 & 15 & 18 & 52 \\

\hline
\multicolumn{16}{l}{\textsc{1.102 Gyr, 3 $M_{Nep}$}}\\
\hline

{\it mm(+3)f5.3n4m}    &15  & $5.09 \pm 0.25$   &4 & 200 &0.04 &30 & 9
    & 0 & 0 & 2 & 21 & 7 & 22 & 32 & 41 \\
    
\tableline

\end{tabular}
\end{center}

Here $N_{pl}$ is the number of planetesimals, $\tau_{dest.}$ the time of destabilization, $\epsilon$ the total planetesimal mass as a fraction of the total planet mass, $N_{sys}$ the number of simulations run in a given suite, $N_{\rm active}$, the number of systems still dynamically evolving after 1.102 Gyr (see the criterion in \S\ref{sec:setup}), $X$ p. the number of systems of multiplicity $X$, and C, D, and E being the numbers of collisions, disruptions, and ejections respectively.

~

NOTE -- By the end of the simulations, the majority of systems with the maximum number of planets remaining do not reach a time of instability; the averages produced here are thence for the other systems. In particular, this skews the 'true' averages for the \textit{d5.6n2m}, \textit{f5.6n1m}, \textit{f5.12n1m} and \textit{f5.12n2m} suites down. Instability (or destabilization) time $\tau_{dest.}$ is measured in years with reference to when the planetesimals were introduced. 

\end{table*}

\section{Results} \label{sec:results}

\begin{figure*}[!tb]
\centering
\includegraphics[width=18cm]{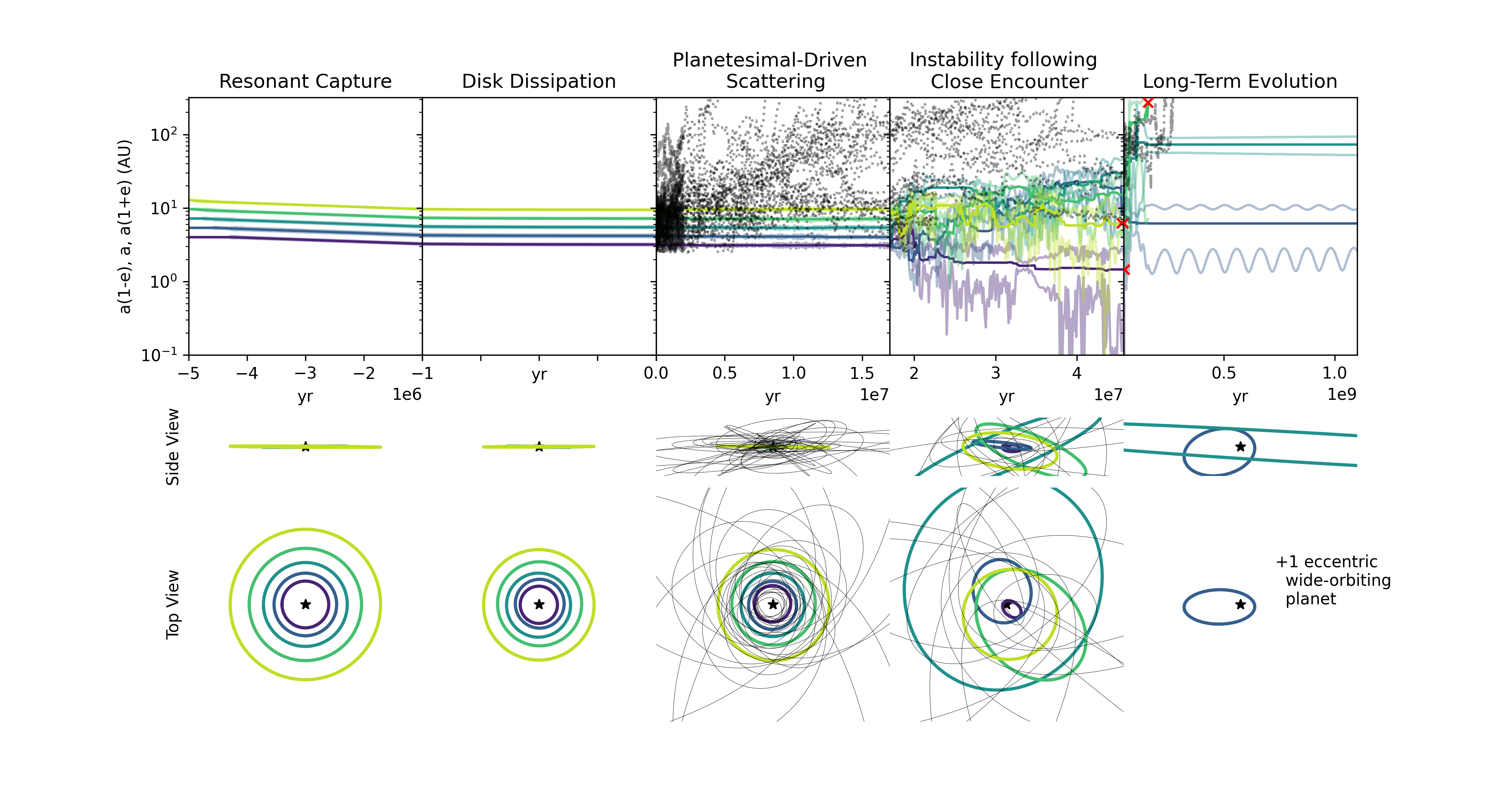}
\caption{Representative orbital evolution from a simulation in the fiducial \textit{f5.6n2m} suite. From left to right, panels show: {\bf i)} capture of cold Neptunes into resonance via disk-driven migration; {\bf ii)} gradual removal of damping representing disk dispersal; {\bf iii)} introduction of planetesimals and scattering of cold Neptunes until the first close encounter; {\bf iv)} onset of global instability, during which one or more planets are lost; and {\bf v)} long-term evolution as the system relaxes to two well-separated planets. The $x$-axis scale differs between panels. 
{\bf Top:} Evolution of semimajor axis, pericenter, and apocenter of each cold Neptune (colored lines, with higher transparency for pericenter and apocenter) and time-averaged orbital separations of planetesimals (dashed black lines). Red markers indicate the loss of a planet by collision, disruption, or ejection. 
{\bf Bottom:} Snapshots of the orbital evolution in side-on and top-down views relative to the initial disk plane. The primary is shown as a black star, planetary orbits as colored ellipses, and planetesimal orbits as thin black ellipses. In the top-down view, orbits are projected onto the disk plane.}
\label{figure:destab}
\end{figure*}

All systems simulated within the \textit{f5} series of simulations are captured into resonance. In particular, the Neptunes are both two-body and three-body resonant at 3:2, with libration amplitudes with 5\textdegree~for two-body resonances and within 3\textdegree ~for three-body resonances. 

We ran a small ($N_{sys}=15$) suite of simulations with no planetesimals---\textit{f5.0n0m}---to confirm the stability of systems produced by our approximated-disk resonant capture. All systems remained stable for in excess of 1 Gyr once the disk was removed in the absence of planetesimals.

Upon introduction of planetesimals, interactions with planetesimals drive the libration amplitude to rapidly increase until the resonant angle begins circulating. The resonance is broken well within the first 100 kyr after the  introduction of planetesimals, but Neptunes retain period ratios approximately that of the ratios of commensurability until the chain destabilizes.

We consider a system to destabilize if, for any pair of planets, the difference between the pericenter of the exterior and apocenter of the interior is less than $CR_H$, for $R_H$ the mutual hill radius and $C$ some coefficient of order unity. For a pair of planets $i$ and $j$ being the interior and exterior planets respectively, this criterion is given by 

\begin{equation}
C R_H \leq a_j (1-e_j) - a_i (1+e_i)
\end{equation}
with 
\begin{equation}
R_H = \frac{(M_ia_i + M_ja_j)}{M_i+M_j} \left( \frac{M_i + M_j}{3M_{\star}} \right)^{1/3}
\end{equation}
the equation of the mutual hill radius \citep{Marchal1982}. For our criterion we set the coefficient $C=1$.

We confirm by visual inspection of over 180 simulations in the \textit{f5} series, including all simulations in the fiducial suite, that the metric marks well the onset of instability in the system. In practice, our destabilization criterion well describes the time of the first close encounter in the system with two exceptions, which we remark upon in the section which immediately follows this one.

\subsection{Schematic Simulations}
\label{subsec:schem_sim}

A summary of the results of our simulations is provided in Table~\ref{table:summary_final}. 
Our N-body simulations neatly divide into three classes characterized by their stability and (if applicable) the nature of their first close encounter:

\begin{figure*}[!ht]
\centering
\includegraphics[width=16cm]{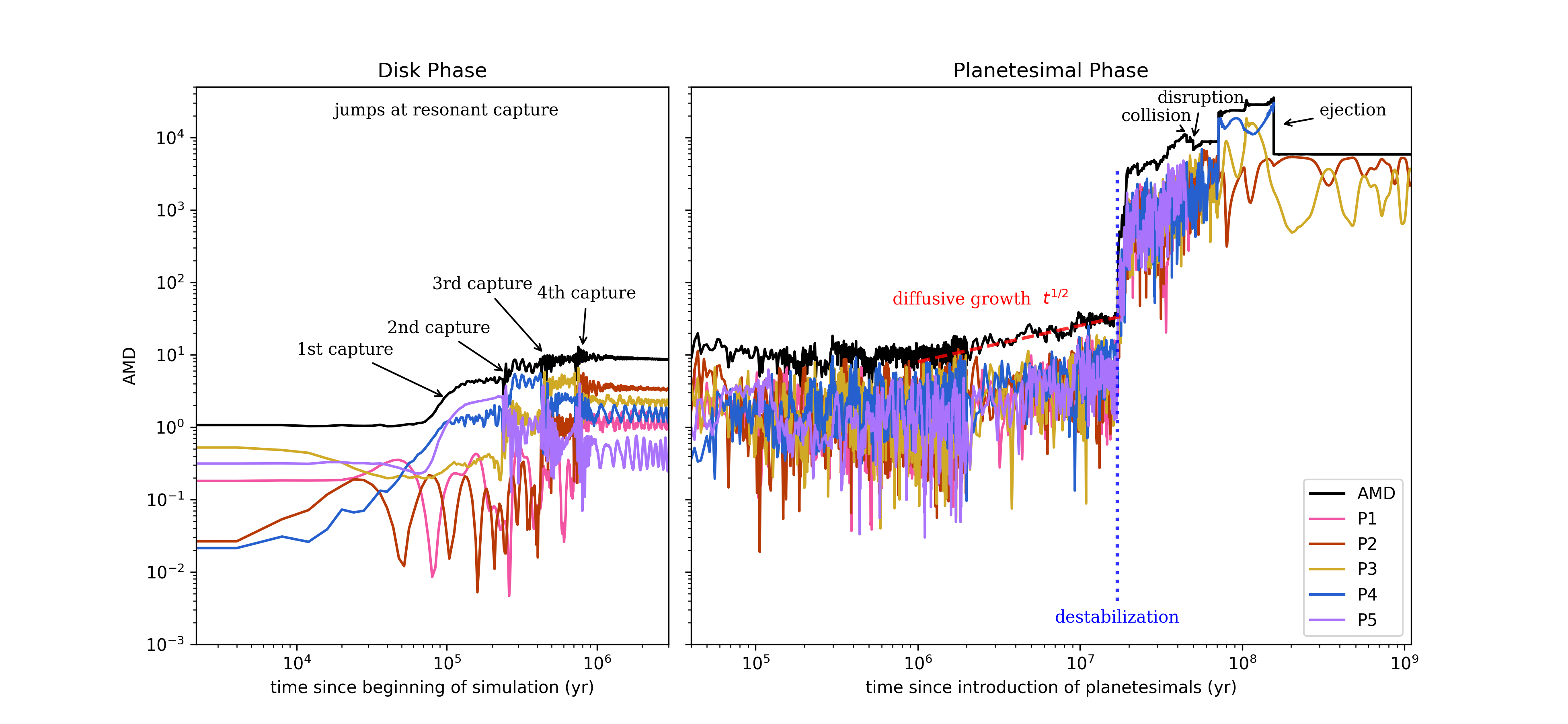}
\caption{Angular momentum deficit (AMD) evolution for the same simulation as in Figure~\ref{figure:destab} from the \textit{f5.6n2m} suite, normalized to its initial value. Colored lines show the contributions to the AMD from individual Neptunes, labeled P1--P5 according to their initial ordering from the primary. The {\bf left} (disk-phase) panel corresponds to the first two panels of Figure~\ref{figure:destab}, while the {\bf right} (planetesimal-phase) panel corresponds to the latter three panels. Annotated features are discussed in \S\ref{subsec:amd}.}
\label{figure:AMD}
\end{figure*}

\begin{description}
\item[Stable]{Stable systems never destabilize and retain all five planets. 
Stable systems are still non-resonant. 
Thirty-two such systems (out of 270) occur in our \textit{f5} series of simulations, not counting the fifteen systems without planetesimals which do not destabilize. 
Among systems with fewer planets (comprising the \textit{d5} and \textit{e5} suites), systems which do not destabilize are more common. 
The \textit{d5} suite contains 17/30 systems which do not destabilize, whereas the \textit{e5} suite contains 2/30 systems which do not destabilize; roughly the same fraction as in the \textit{f5} series of simulations. 
Systems which retain all their planets are not guaranteed to be stable; they may have destabilized but have yet to experience a close encounter resulting in a collision, disruption, or ejection.}

\item[Single CE]{Single close encounter systems are characterized by a single close encounter during their evolution. Prototypically, the two innermost planets undergo a close encounter in which they collide, and the rest of the system is non-resonant but stable.\footnote{Single CE systems are not marked as destabilized because the encounter occurs between snapshots, and the stable state following the encounter does not meet the criterion for instability.} 
Resultant four-planet systems experience no close encounters for the remainder of the integration. 
One such system occurs in our \textit{f5} simulations (simulation 4 of \textit{f5.12n2m}) and one occurs in the \textit{e5} series (simulation 6 of \textit{e5.6n2m}). }

\item[Multiple CE]{Multiple close encounter systems are characterized by an initial single close encounter followed quickly by close encounters with other planets. Numerous close encounters rapidly cause all planets' orbits to vary dramatically, leading to a global dynamical instability, subsequently settling to a state with few planets remaining. The simulation presented in Figure \ref{figure:destab} is a multiple close encounter system. Multiple close encounter systems account for the majority of our \textit{f5} simulations.}
\end{description}

Although only two single close encounter systems occur in our simulations, the category is illustrative of how close encounters result in instability. 
In a system just prior to its first close encounter, a pair of planets' eccentricities are sufficiently high to permit a physical separation of less than a mutual hill radius. 
When the orbits are aligned such that this occurs, planets then approach each other in semimajor axis, departing from their original orbits, eventually resulting in a close encounter. 
If the close encounter is a scattering, the dramatic changes in orbital parameters almost always result in one of the involved planets experiencing a close encounter with another planet due to the tightly-packed nature of the resonant chain\footnote{As a counterexample, in one preliminary simulation the two planets involved in the first close encounter traded semimajor axes, otherwise remaining relatively stable; global instability occurred only after a second close encounter involving a different pair.}---thus instability follows. 
If the encounter is a collision, the merged planet's parameters are unlikely to permit an encounter with another planet. 
No further close encounters are generated, thus producing a single CE system. 

The typical simulation is a multiple close encounter system in which the following occur. 
First, subsequent to the disk phase, planetesimals are introduced into the simulations. 
Second, as the system evolves, planetesimal-induced perturbations break resonances amongst the planets. 
Third, the now non-resonant chain of planets collide with and scatter the planetesimals. 
Fourth, as most planetesimals are being scattered, rising eccentricities result in the first close encounter among the planets. 
Fifth, the system destabilizes after its first close encounter. 
Sixth, once destabilized, losses of planets occur through collisions, disruptions, and ejections, over a further Gyr of integration. 
The most common end state for our simulations is that of two to three eccentric Neptunes remaining in the system.  

\subsection{AMD Evolution}
\label{subsec:amd}

We parameterize instability in the system with a quantity denoted the \textit{angular momentum deficit}, or AMD, which is given by \citet{Laskar1997} as 
\begin{equation}
    C_{\rm AMD} = \sum_{\textrm{planets}~k} \Lambda_k \left( 1 - \sqrt{1-e_k^2} \cos{i_k} \right)
\end{equation}
with  
\begin{equation}
    \Lambda_k = \frac{m_k M_*}{m_k + M_*} \sqrt{G(m_k + M_*) a_k}
\end {equation}
and subscripted orbital parameters corresponding to the orbital parameters of a planet $k$, with the sum being taken over all planets. The AMD represents the excitation of planetary orbits from exactly circular and coplanar orbits; in that state, the AMD has a value of 0. 
Losses of planetary bodies from the system, whether by collisions, disruptions, or ejections, result in a reduction in the AMD~\citep{Laskar2017}. 
Resonant interactions and close encounters can result in an increase in the AMD~\citep{Laskar2017,Murphy2022,HaddenWu2025}.
Under secular evolution, angular momentum deficit is conserved~\citep{Laskar1997}.

Figure~\ref{figure:AMD} displays the evolution of the angular momentum deficit in a characteristic system.
In this figure, the contribution to the AMD from planetesimals is ignored, and the AMD is normalized to the initial conditions of the system's planets.
During the disk phase, the angular momentum deficit increases substantially with each resonant capture.
At later times, as systems approach destabilization, the AMD grows roughly in proportion to the square root of the time elapsed\footnote{Unlike \citet{HaddenWu2025}, in our simulations such diffusion-like behavior occurs prior to destabilization, and AMD evolution subsequent to destabilization is much more chaotic.}.

We remark qualitatively that the AMD enters an approximate state of equipartition between planets during this phase of diffusion-like growth. 
The visual signature of this approximate equipartition in Figure~\ref{figure:AMD} is the approximate overlapping placement of all planets' median AMD about a half order of magnitude below the total AMD, indicating that no one planet dominates nor contributes substantially less to the AMD budget than the others.

Once a system destabilizes, the AMD rapidly increases by up to several orders of magnitude and remains elevated, decreasing slightly when planets are lost. 
In a system of approximately equal-massed planets, when a planet is forced into the outer reaches of the system, it tends to dominate the angular momentum budget due to the increased $\Lambda_k$ factor; correspondingly ejections tend to remove much more AMD from the system than disruptions or collisions which typically occur more frequently among inner planets.

In Figure~\ref{figure:AMD}, this difference is exemplified in the original fourth planet from the primary (P4), which from 70-160 Myr largely dominates the AMD. When ejected, the system's AMD drops by about 80\%. 
Upon settling to a long-term stable two-planet state with significant orbital separation between the planets, AMD tends to be conserved, as the system can then be well described secularly. 
For destabilized systems, we find that AMD rises by three to four orders of magnitude from the original non-resonant in-disk configuration after 1 Gyr.

\begin{figure*}[!htb]
\centering
\includegraphics[width=16cm]{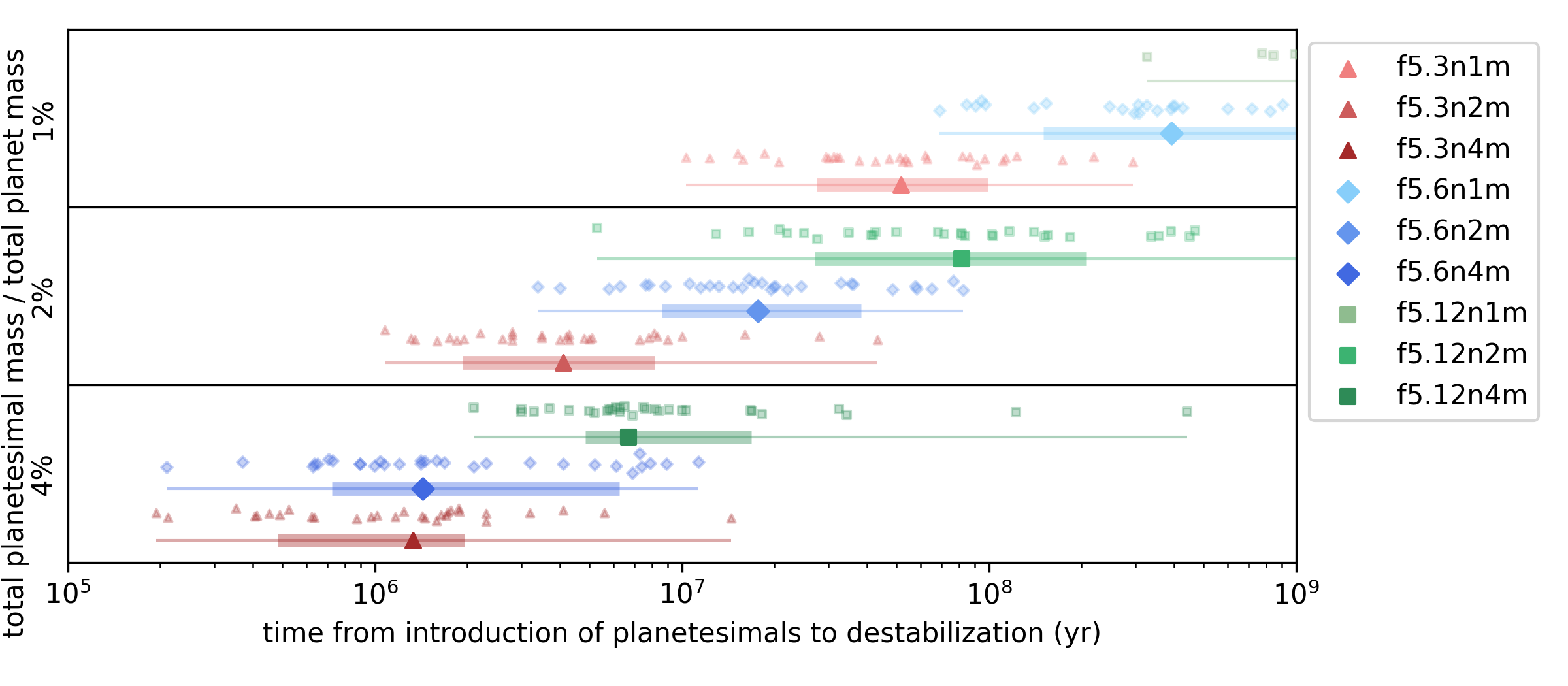}
\caption{Destabilization times for five-planet systems, defined as the time of the first close encounter between any pair of planets. Colors and symbols distinguish between simulation suites. Time is measured from the introduction of planetesimals. Each small symbol represents an individual system, while large symbols indicate the median destabilization time. Distributions are shown as lines: thin lines trace the full distribution, and thick lines mark the range between the 20th and 80th percentiles. Some systems do not destabilize within 1.1 Gyr; accordingly, distributions for those suites extend beyond the edge of the figure.}
\label{figure:destabilization_times}
\end{figure*}

\subsection{Destabilization Times}
\label{subsec:destab_t}

In our \textit{f5} suite of simulations, we find that destabilization times are both highly sensitive to planetesimal disk conditions and exhibit broad distributions.
Destabilization times depend both on the total mass of planetesimals and the granularity of the planetesimals and range from 1-1000 Myr. 
Grainy planetesimal systems, with more mass per individual grain for a fixed proportion of the planets' masses, tend to destabilize faster than smooth planetesimal systems. 
Based on the limited selection of suites, the granularity is of secondary importance to the total mass of the planetesimals. 
Among systems which destabilize, times for destabilization vary significantly, differing by up to two orders of magnitude between simulations with the same planetesimal disk parameters. 
The distribution of destabilization times is provided in Figure \ref{figure:destabilization_times}. 
Means and standard deviations of the logarithm of destabilization time are reported in Table \ref{table:summary_final}. Even for systems with the same planetesimal disk parameters, destabilization times can differ dramatically. Moreover, the distributions are sufficiently broad that destabilization time alone cannot be used to infer mass and granular properties of a planetesimal disks.

Some systems despite the perturbations induced by planetesimals do not destabilize. Of the 270 \textit{f5} simulations run and completed within the computational time allotted, thirty-three systems do not globally destabilize; all non-destabilizing systems are in configurations with longer log-mean destabilization times than 100 Myr, and about 75\% are within the \textit{f5.12n1m} suite.

\begin{figure}[!tb]
\centering
    \includegraphics[width=8cm]{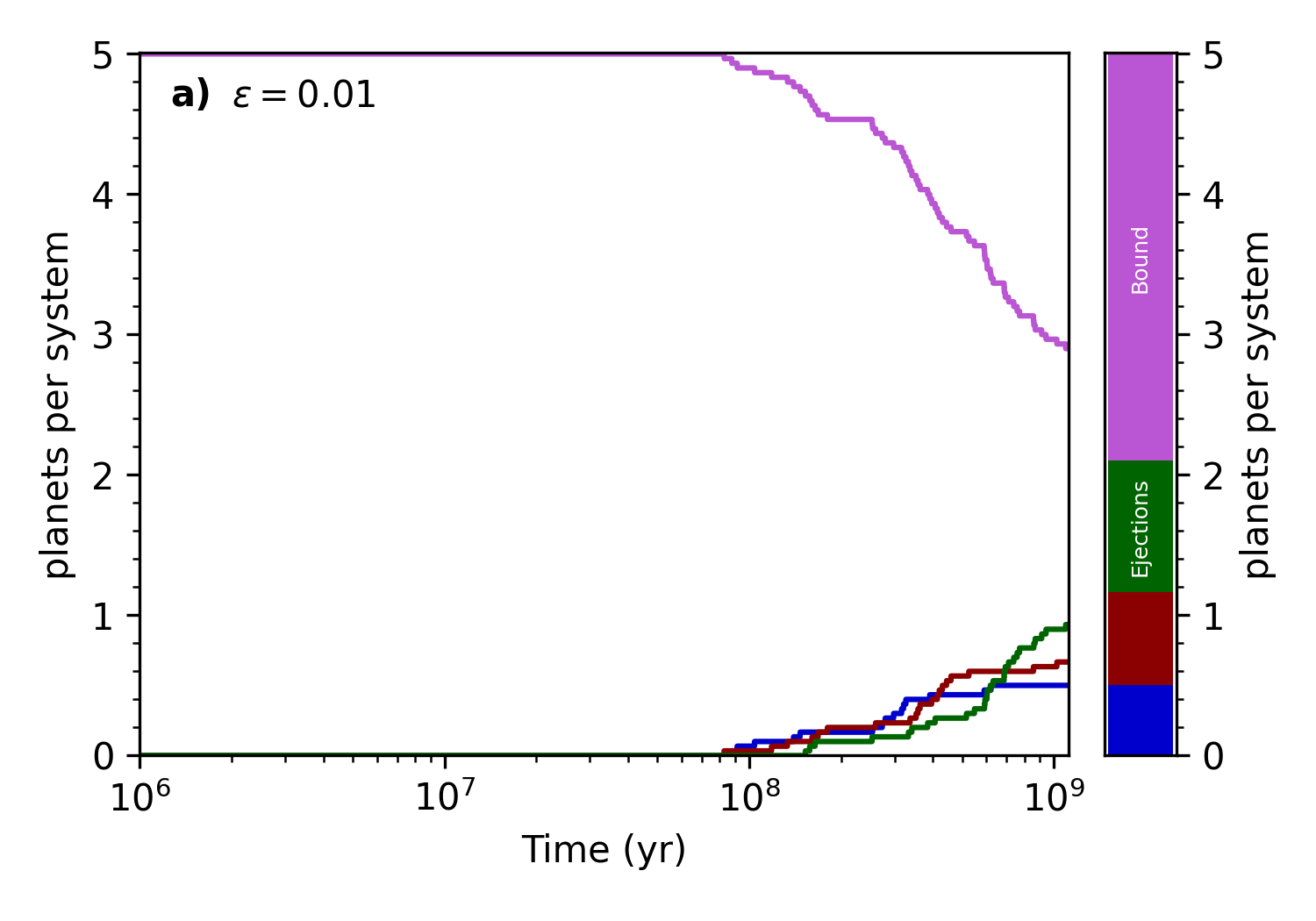}
    \includegraphics[width=8cm]{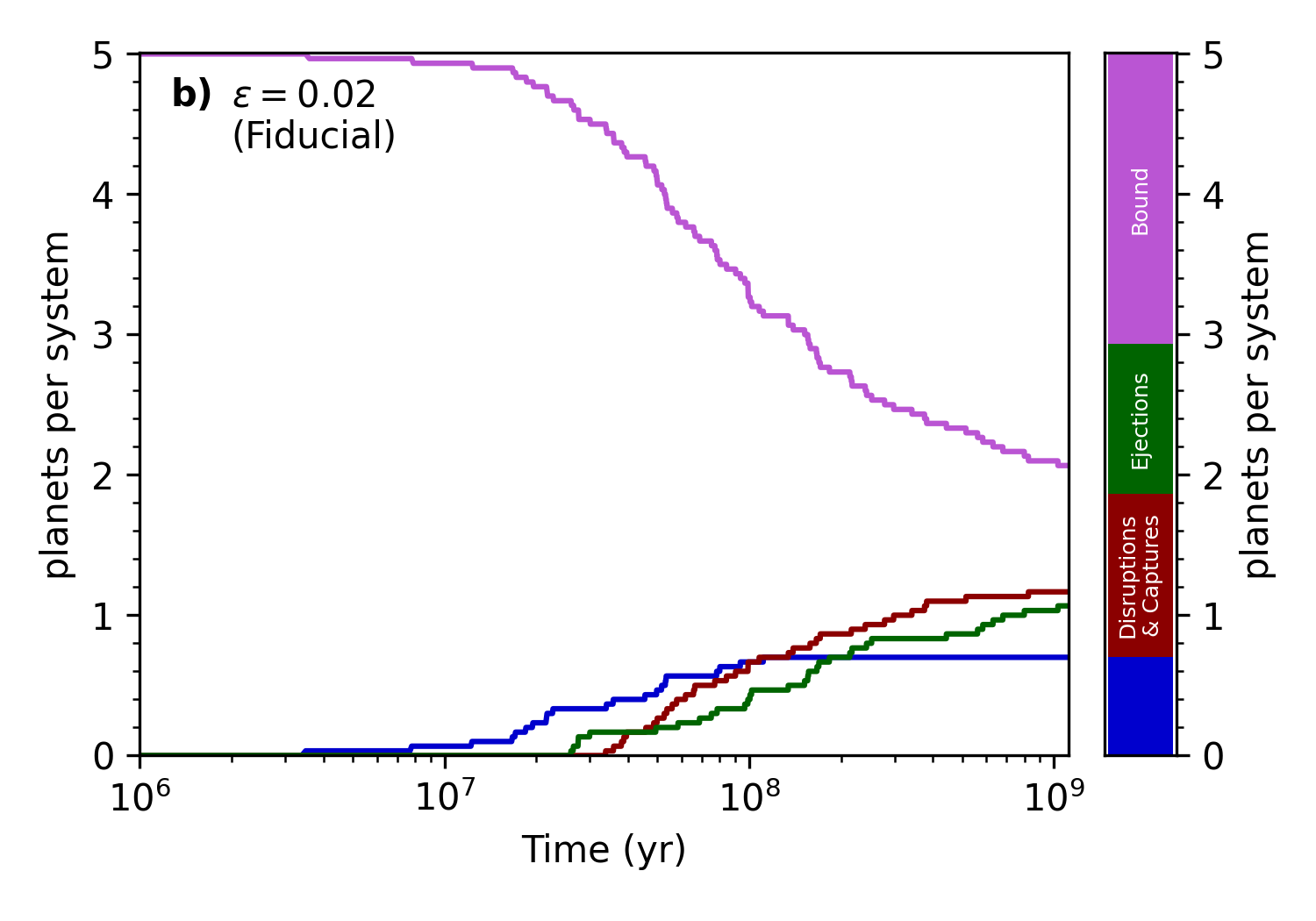}
    \includegraphics[width=8cm]{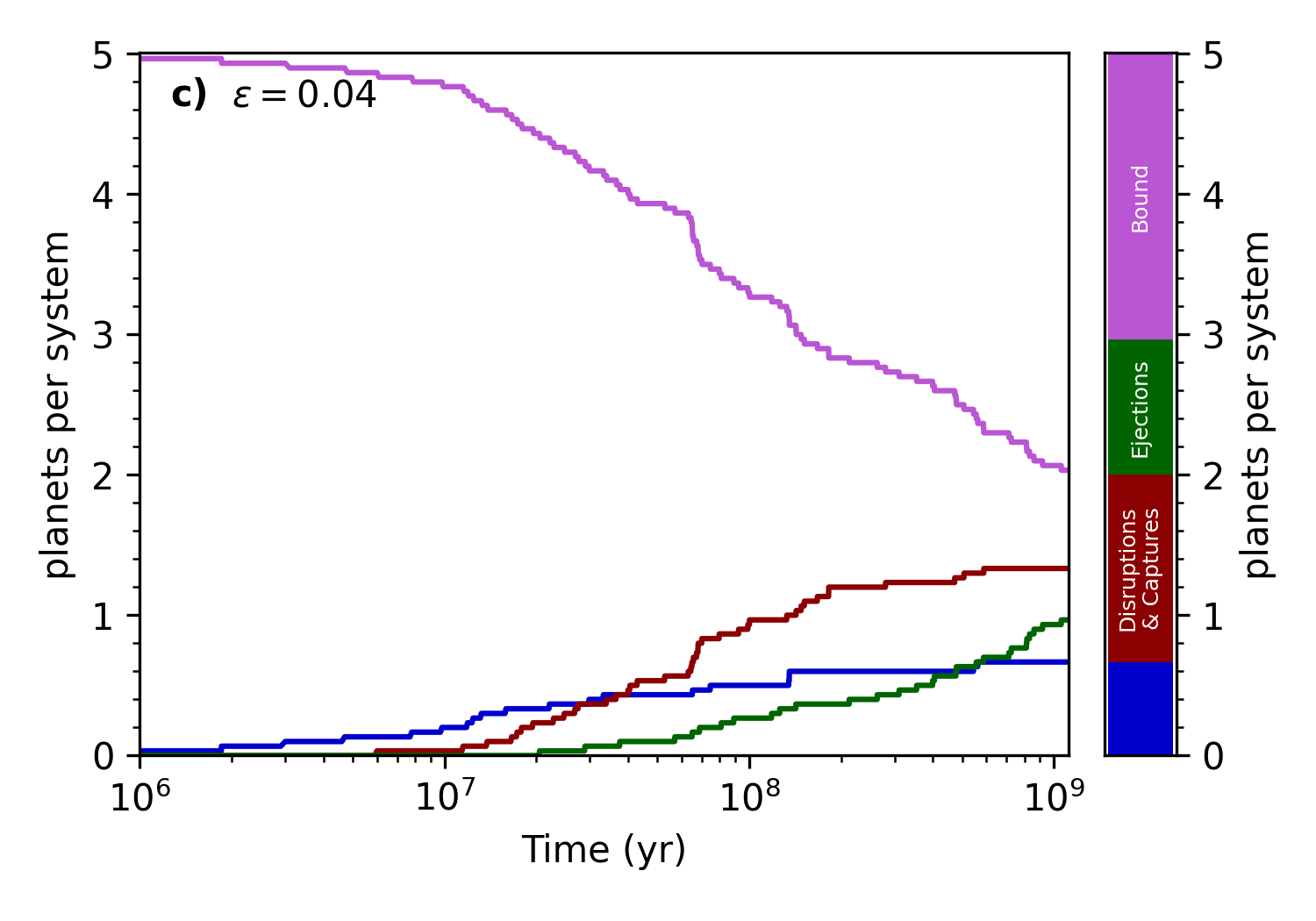}
    \includegraphics[width=4cm]{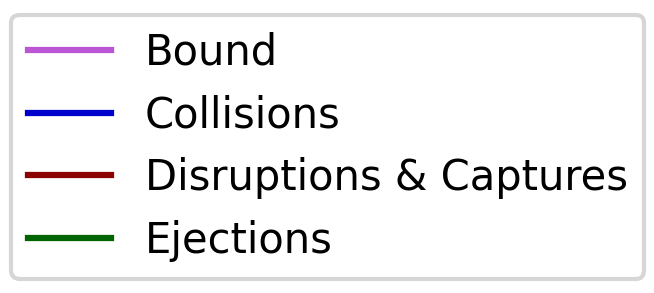}

~

~

\caption{Average number of planets per system in each outcome as a function of time across the simulation suite. Colors denote the class of state, while the color bar on the right indicates the average number of planets per system in the corresponding end state. Panels show suites of thirty planetesimals with different planetesimal masses (values of $\epsilon$), increasing from top to bottom.
}
\label{fig:cde}
\end{figure}

\begin{figure}[!tb]
\centering
\includegraphics[width=8cm]{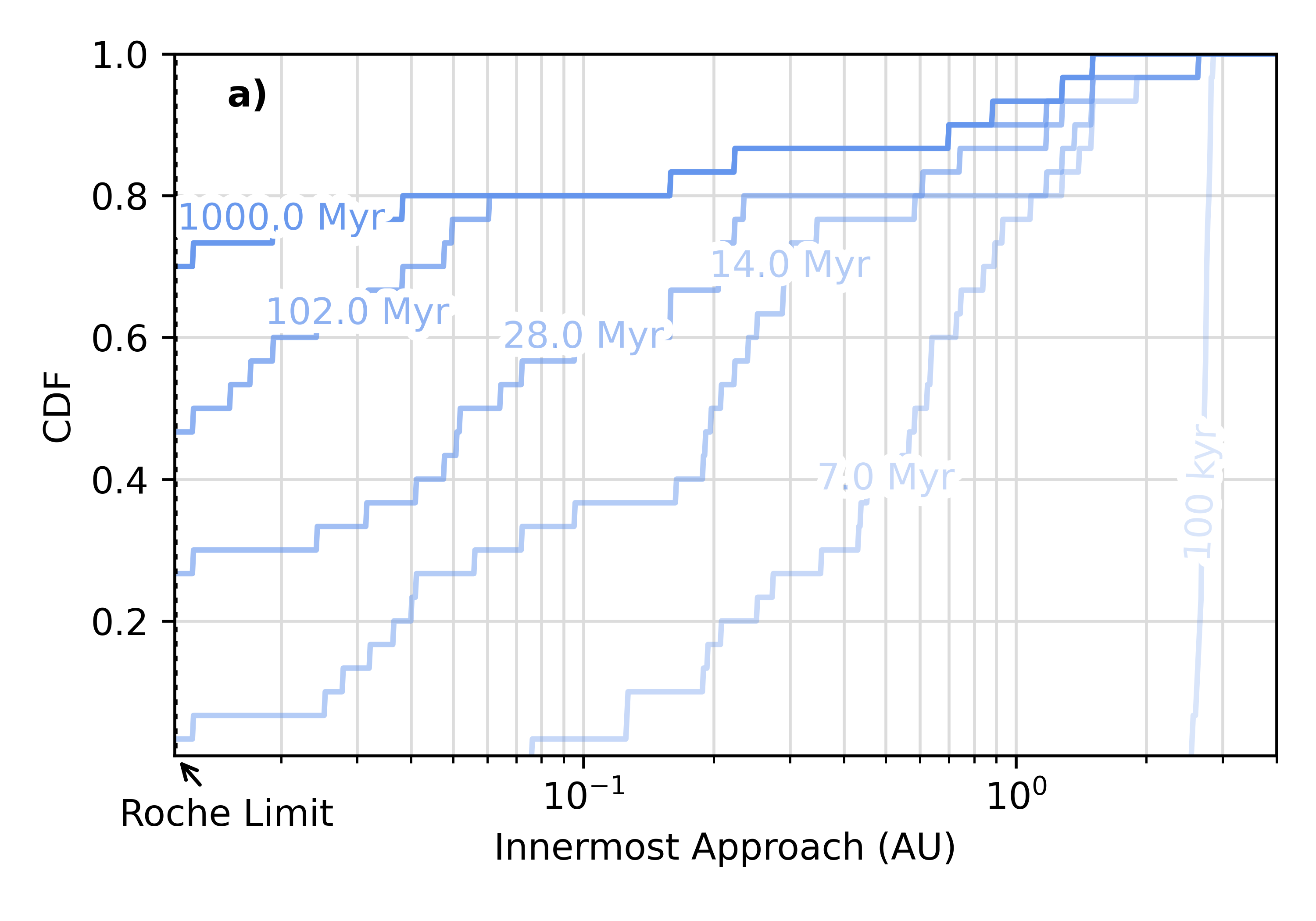}
\includegraphics[width=8cm]{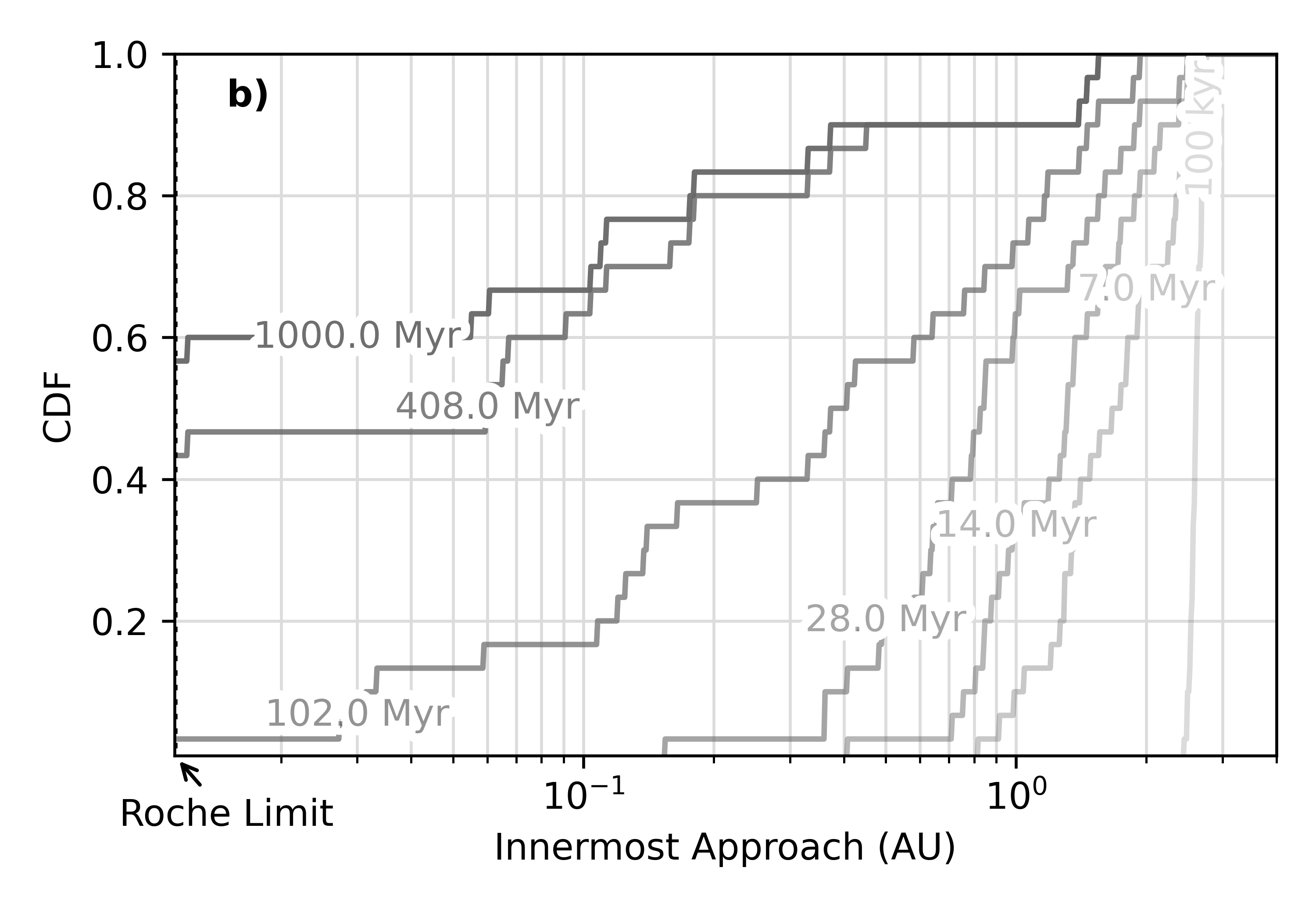}
\caption{Cumulative distribution functions of the minimum pericenter distance reached by any planet in a given system at selected times. Times are measured in years since system destabilization and are ordered from earliest to latest by increasing opacity.
Panel {\bf a)} corresponds to the fiducial \textit{f5.6n2m} suite, and panel {\bf b)} corresponds to the sub-Neptune \textit{mm(-3)f5.3n4m} suite ($1/3$ Neptune mass).
The innermost boundary marks the disruption boundary (Roche limit).
}
\label{figure:close_approach}
\end{figure}

\subsection{Collisions, Disruptions, \& Ejections}
\label{subsec:cde}

From purely Safronov arguments, the inner planets of an equal-massed set are more likely to collide than the outermost, as the orbital velocity is greater and thus the Safronov number (Eqn.~\ref{eqn:safronov}) lower. 
Close encounters take place on timescales which scale positively with orbital period. 
We therefore expect that collisions are more likely to happen before ejections in these chains. 

The timing of disruptions is more nuanced. Let us treat disruptions as the result of scatterings from close encounters amid the innermost planets. Inner planets in a set are more likely to be disrupted, as they are closer to the star and thus their eccentricities need be excited to less extreme values. Combined with the reduced frequency of scatterings among the innermost planets in a chain, we expect that the timing of disruptions is later than that of collisions.

These trends are observed in Figure~\ref{fig:cde}, which shows the average number of bound (pink) Neptunes which remain in the system and collided (blue), disrupted (red), and ejected (green) Neptunes which are lost from the system, and the timing at which those losses occur. 
On average, collisions occur before disruptions which occur before ejections. This is most visible in panel \ref{fig:cde}c, corresponding to the \textit{f5.6n4m} suite.
The dependence of destabilization time on planetesimal disk mass is also apparent, with lower disk mass systems accumulating planetary losses much later than higher disk mass systems. 
In panels \ref{fig:cde}b and \ref{fig:cde}c collisions tend to level off at around 100 Myr while disruptions and ejections are still prevalent well after the cessation of collisions. 
This effect is less prominent in panel \ref{fig:cde}a due to a longer median destabilization time and thus presence of systems which have yet to destabilize. 

In broad strokes, across the \textit{f5} series the number of planets disrupted or ejected are approximately the same, with disruptions (1.1/system) and ejections (1.1/system) being more common than collisions (0.6/system). 
The mean number of planets left bound in a previously destabilized system is about twice that the number ejected or disrupted (2.2/system). 

We find that the high rate of disruptions tends to be driven by the mass of the planetesimals at the high end of the mass spectrum.
In our fiducial simulations, the ratio between scattering types $\rm{D/E} = 35/32 \sim 1.1$ indicates that disruptions and ejections are approximately equally likely (see Table~\ref{table:summary_final}). 
Between suites of \textit{f5} simulations, the ratio between scattering types $\rm{D/E}$ tends to vary with the planetesimal granularity, with values of 1.4 for more granular (3 planetesimals/planet) scattering and 0.4 for more smooth (12 planetesimals/planet) scattering. 
While the $\rm{D/E}$ ratio is substantially lower in the smoother than the granular case, the magnitude of the difference may be influenced by complete lack of collisions and near complete lack of disruptions in the $\textit{f5.12n1m}$ suite, possibly as a result of the small sample of destabilized systems in that suite. 
Additional work is necessary to accurately characterize this dependence.

We emphasize that D/E numbers may be biased in favor of ejections on account of numerical error. In 10\% (N=3/30) of the simulations in our fiducial suite, a planet with a very close approach to the star (within 0.03 AU) is ejected from the system, and a jump in energy comparable to collision with the star is recorded. These suspected `spurious' ejections may be reflective of disruptions. A better treatment more accurately accounting not only for disruptions but also tidal captures will be part of a future work.

Other observed trends are not as strong as that of D/E with granularity; no clear trend in planetesimal disk parameters emerges with respect to other branching ratios, nor between planetesimal disk mass and the ratio of disruptions to ejections.

\begin{figure*}[!hbt]
\centering
\includegraphics[width=17cm]{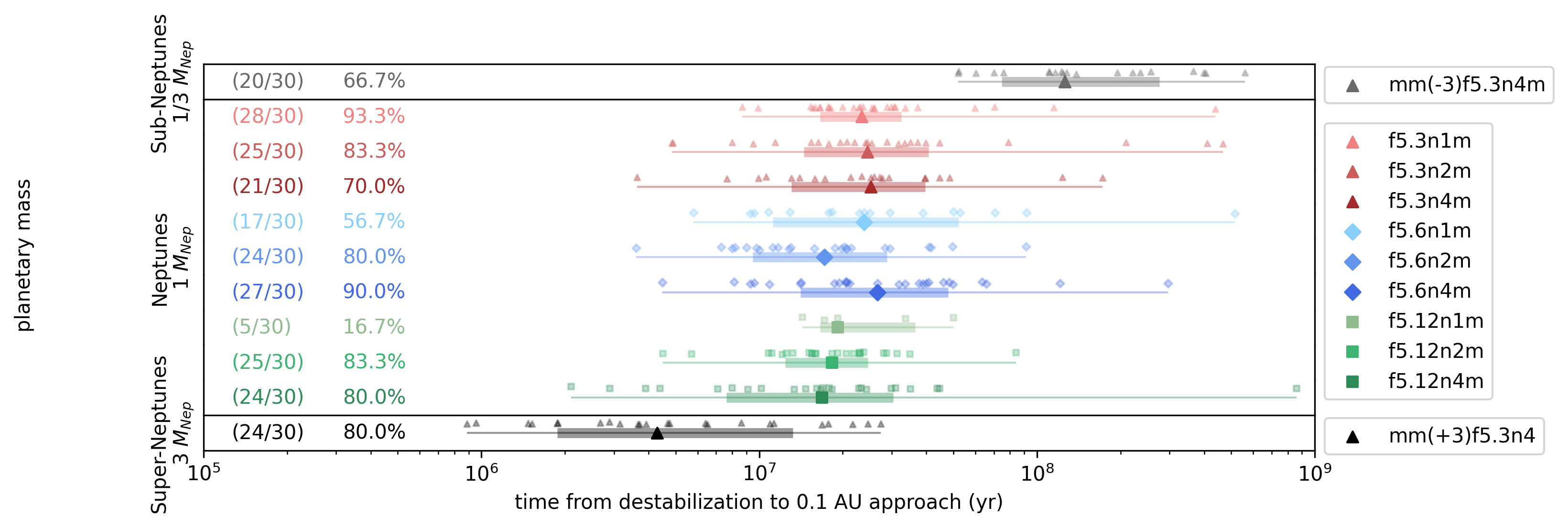}
\caption{Times of first approach within 0.1 au for five-planet systems. Colors and symbols distinguish between simulation suites, with the \textit{f5} series and other series shown in separate panels. Only systems that both destabilize and reach a first approach within 1.1 Gyr are included; the fraction of such systems in each suite is indicated by the ratios on the left. Time is measured from destabilization. Each small symbol represents an individual system, while large symbols indicate the median approach time. Distributions are shown as in Figure~\ref{figure:destabilization_times}.}
\label{figure:close_approach_times}
\end{figure*}

\subsection{Inner System Approaches}
\label{subsec:isa}
After destabilization of a system, the evolutions of orbits in a system are chaotic. Interacting Neptunes rapidly explore the parts of the system interior to their original configuration. In most systems (N=217/270) at least one Neptune achieves an eccentricity sufficiently high to send its pericenter below 0.3 au during its orbital evolution. These Neptunes are typically inclined, with median inclinations between 30-50\textdegree~depending on the suite of simulations. Systems containing a Neptune with eccentricity sufficiently high to drift below 0.3 au in pericenter often contain at least one Neptune whose pericenter drifts below 0.1 au (N=196/270). Closer approaches to the star are accompanied by slightly higher median inclinations; reaching to inclinations of between 40-60\textdegree~for pericenter passages below 0.1 au. 

For the fiducial case and the sub-Neptune case, the innermost expected approach achieved by a planet in a given system by a particular time after destabilization are given in Figure~\ref{figure:close_approach} as cumulative distribution functions. 
Most systems explore relatively constrained areas within the first few 100 kyr after destabilization, with that region widening as eccentricities excite with the scattering of planetesimals. 
The shapes of the CDFs in Figure~\ref{figure:close_approach} imply on average a slow exploration of the inner system; the median pericenter distance gradually approaches the Roche limit.
In the fiducial case, between about 14 Myr and 100 Myr, approach distances indicate that systems begin to experience disruptions. 
In the sub-Neptune case, approach distances indicate that the interior regions of systems are explored more slowly, with disruptions experienced at on the order of 100s of Myr.

\begin{figure*}[!thpb]
\centering
\includegraphics[width=5.4cm]{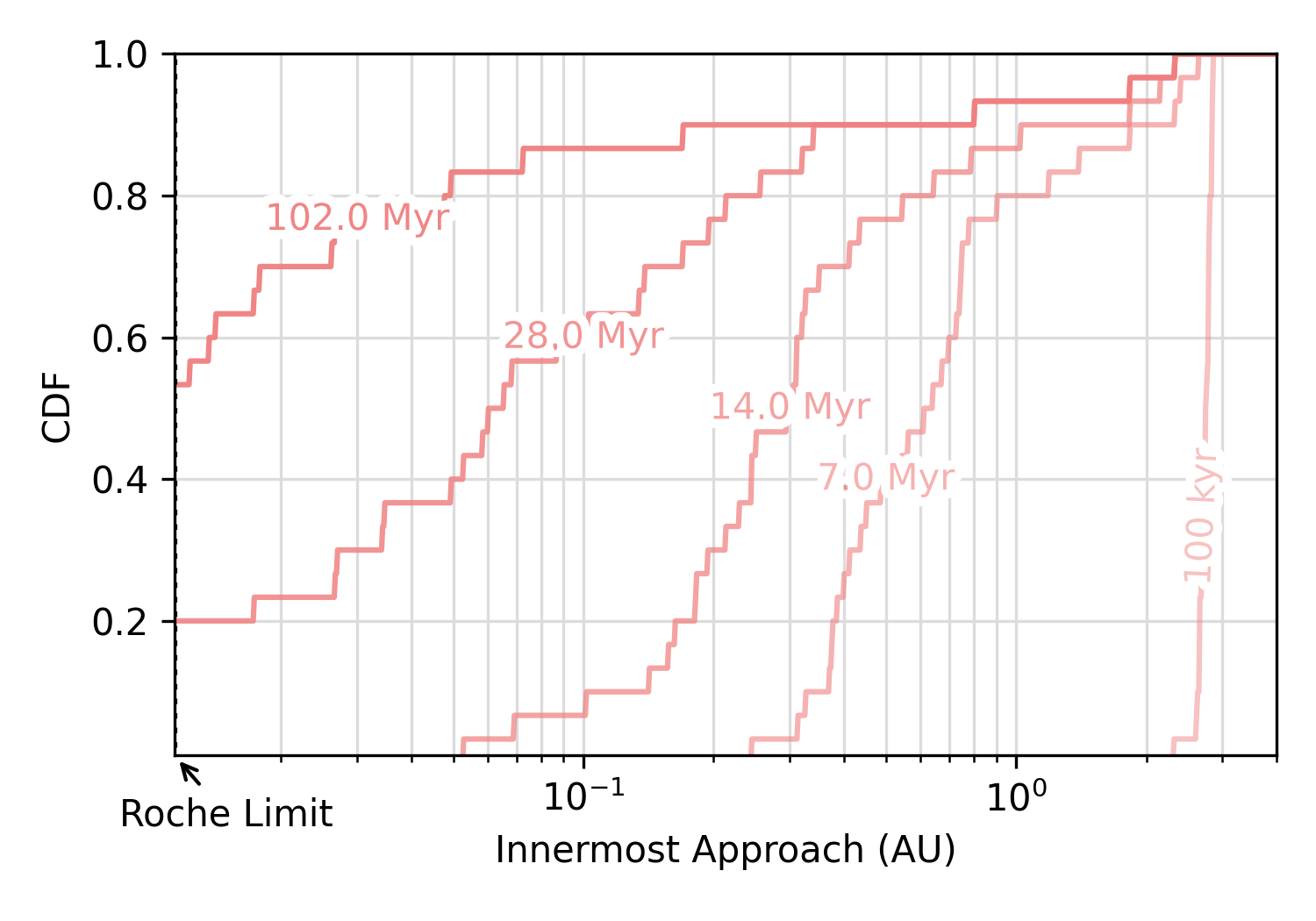}
\includegraphics[width=5.4cm]{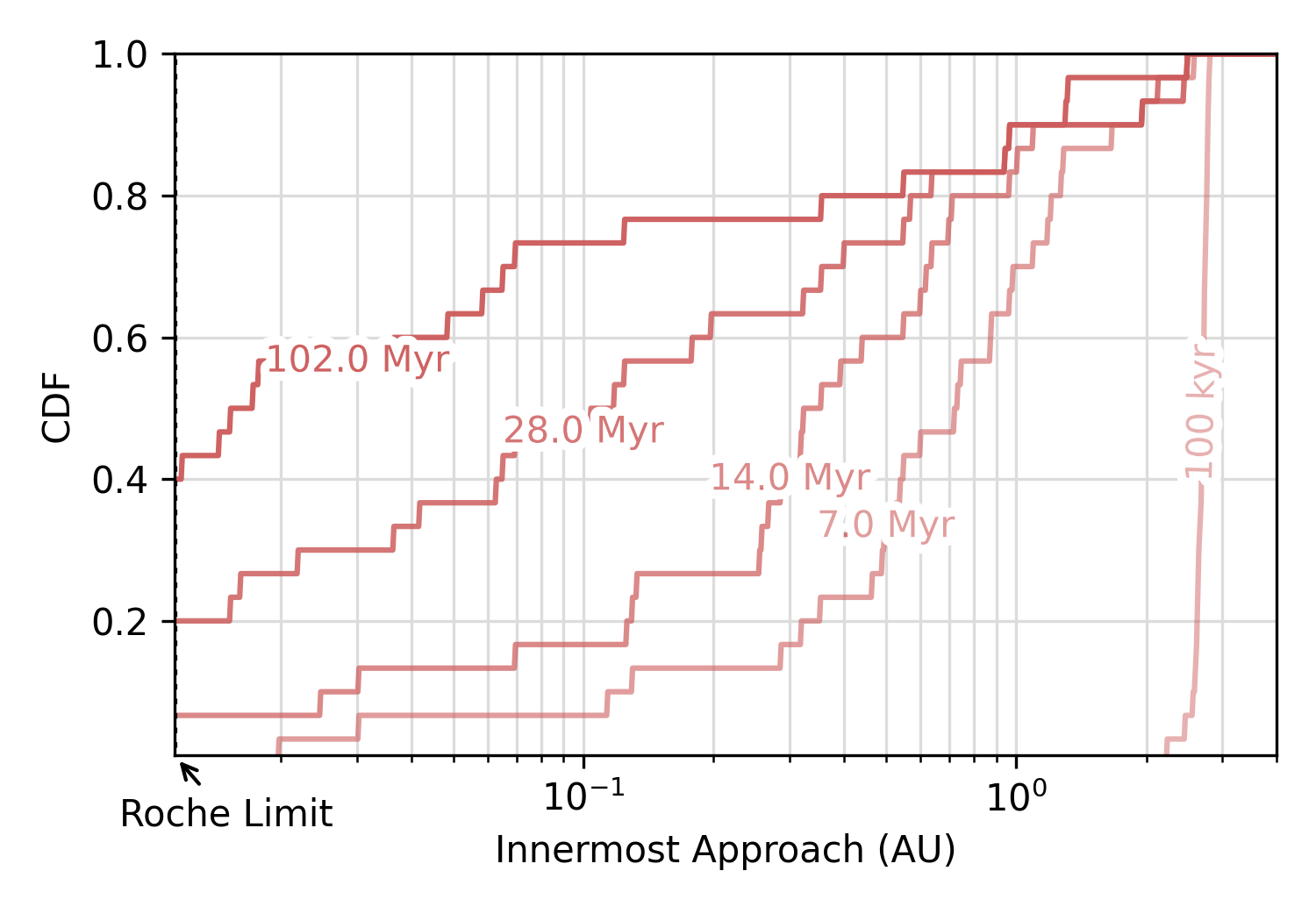}
\includegraphics[width=5.4cm]{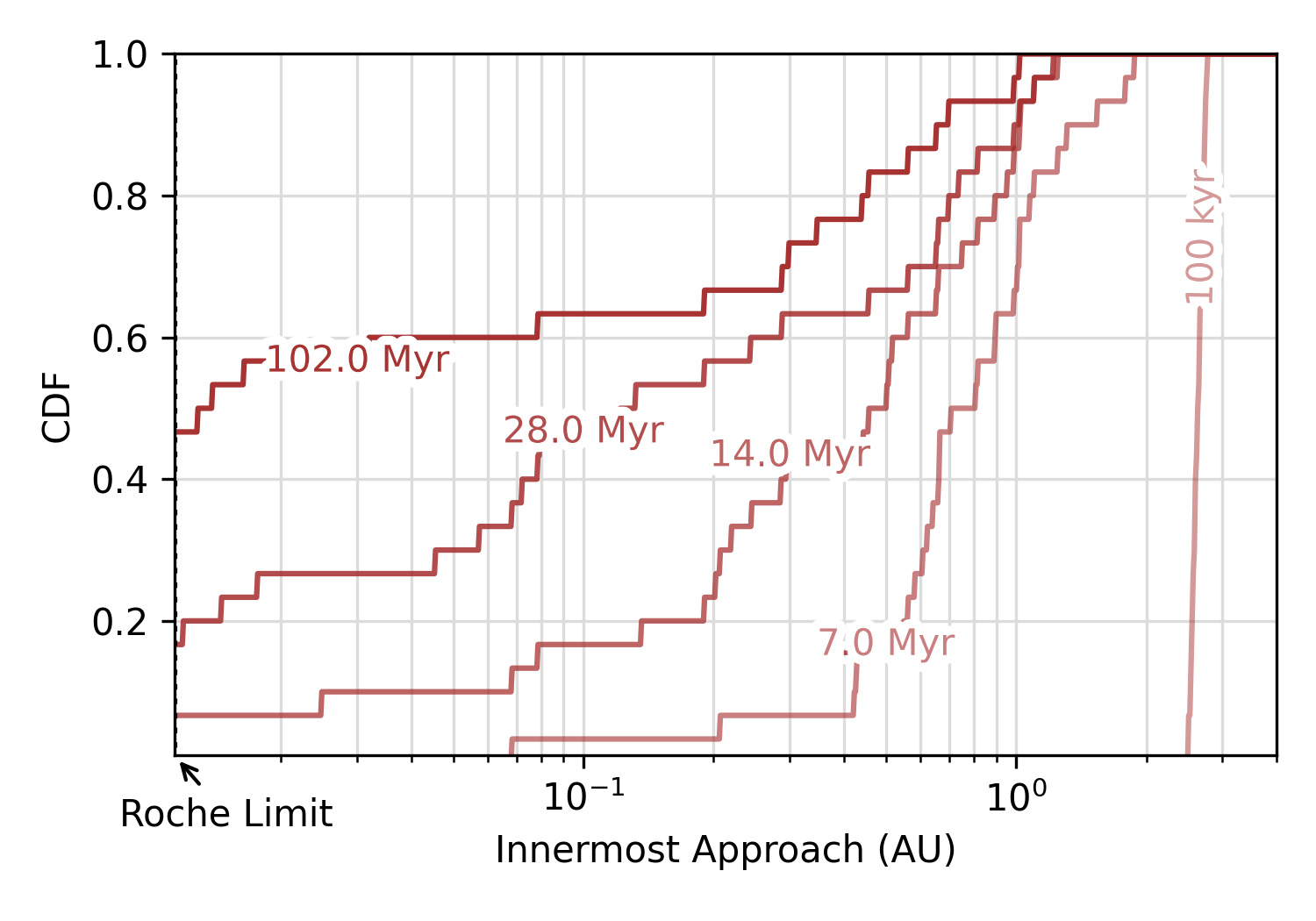}
\includegraphics[width=5.4cm]{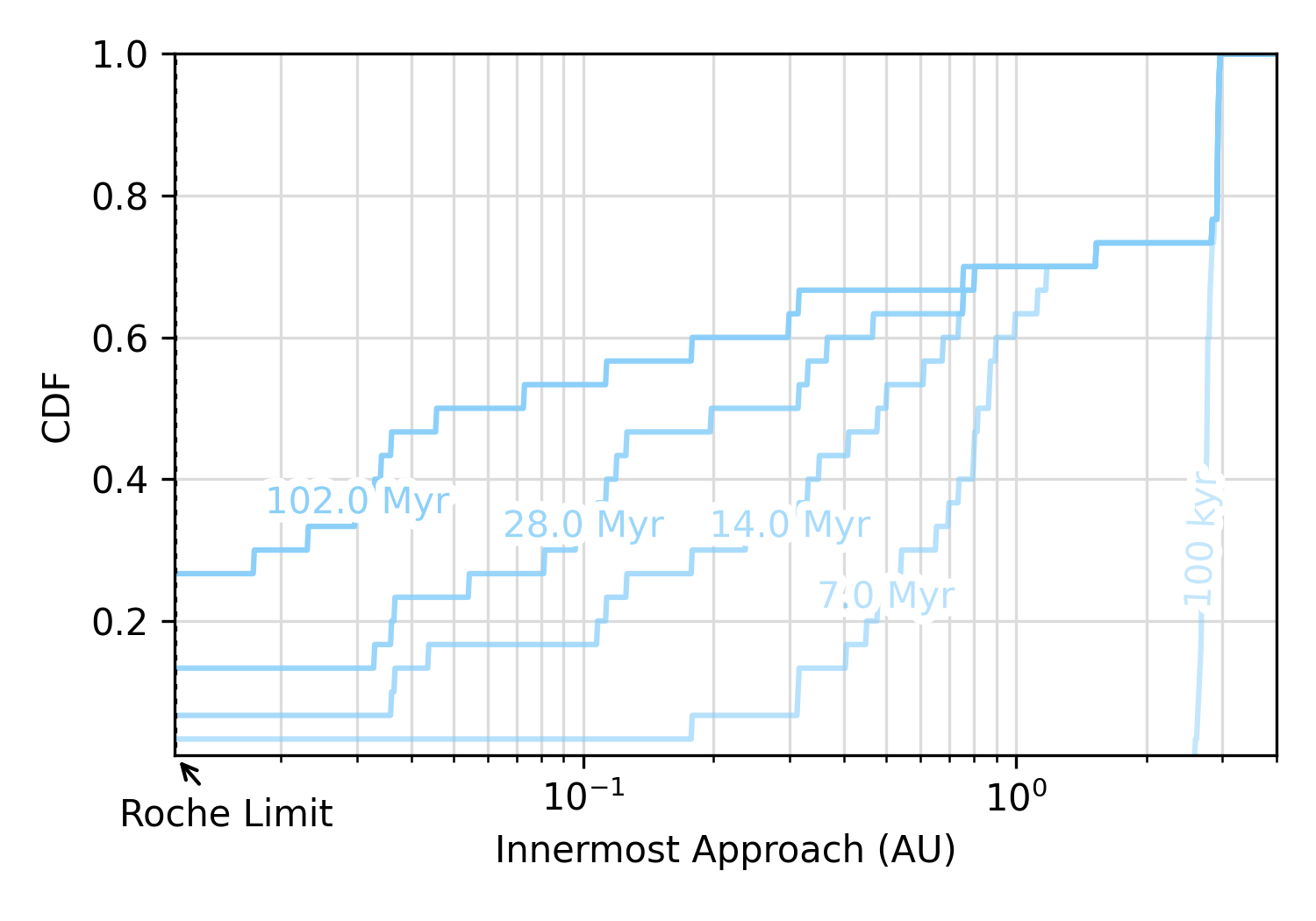}
\includegraphics[width=5.4cm]{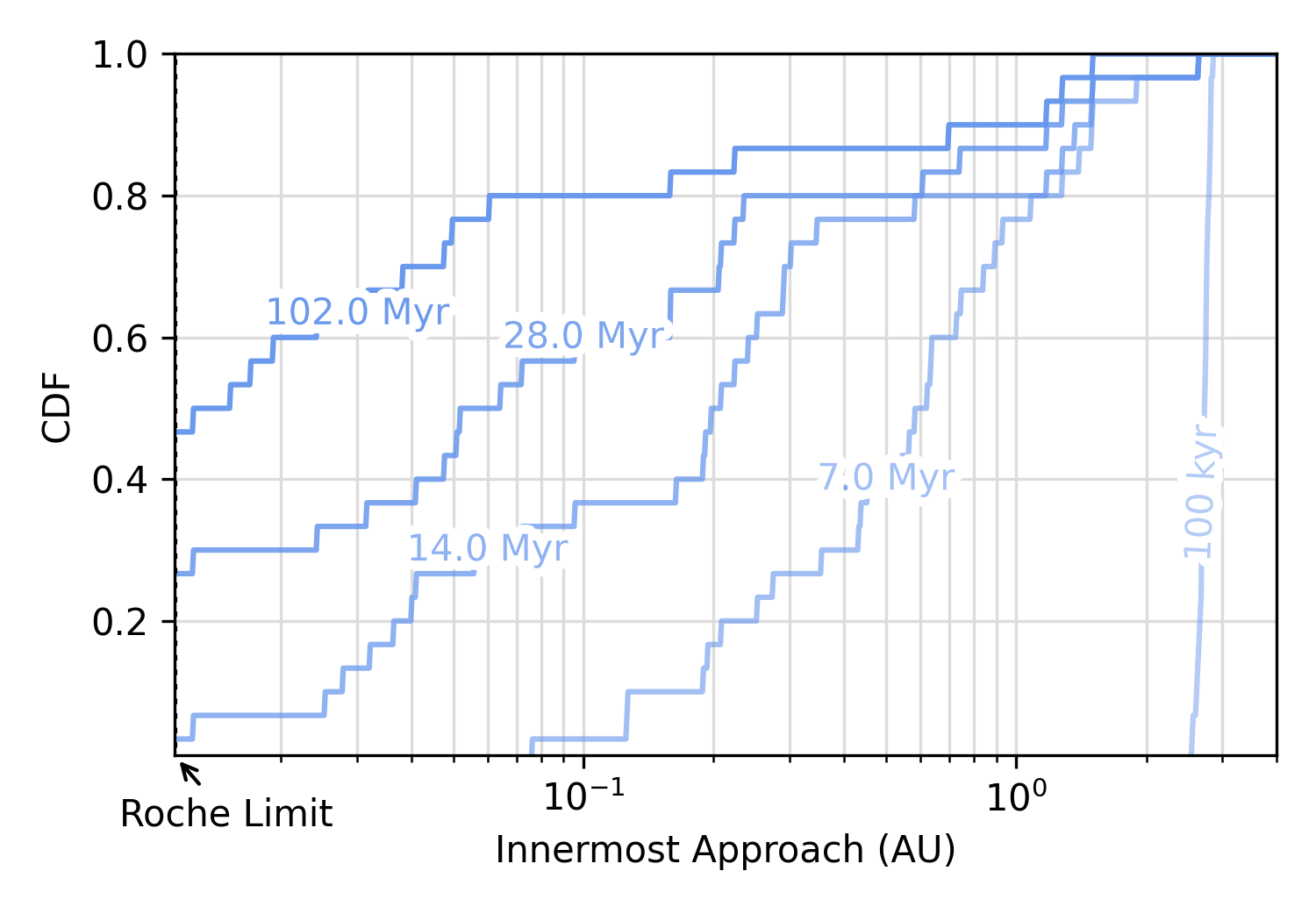}
\includegraphics[width=5.4cm]{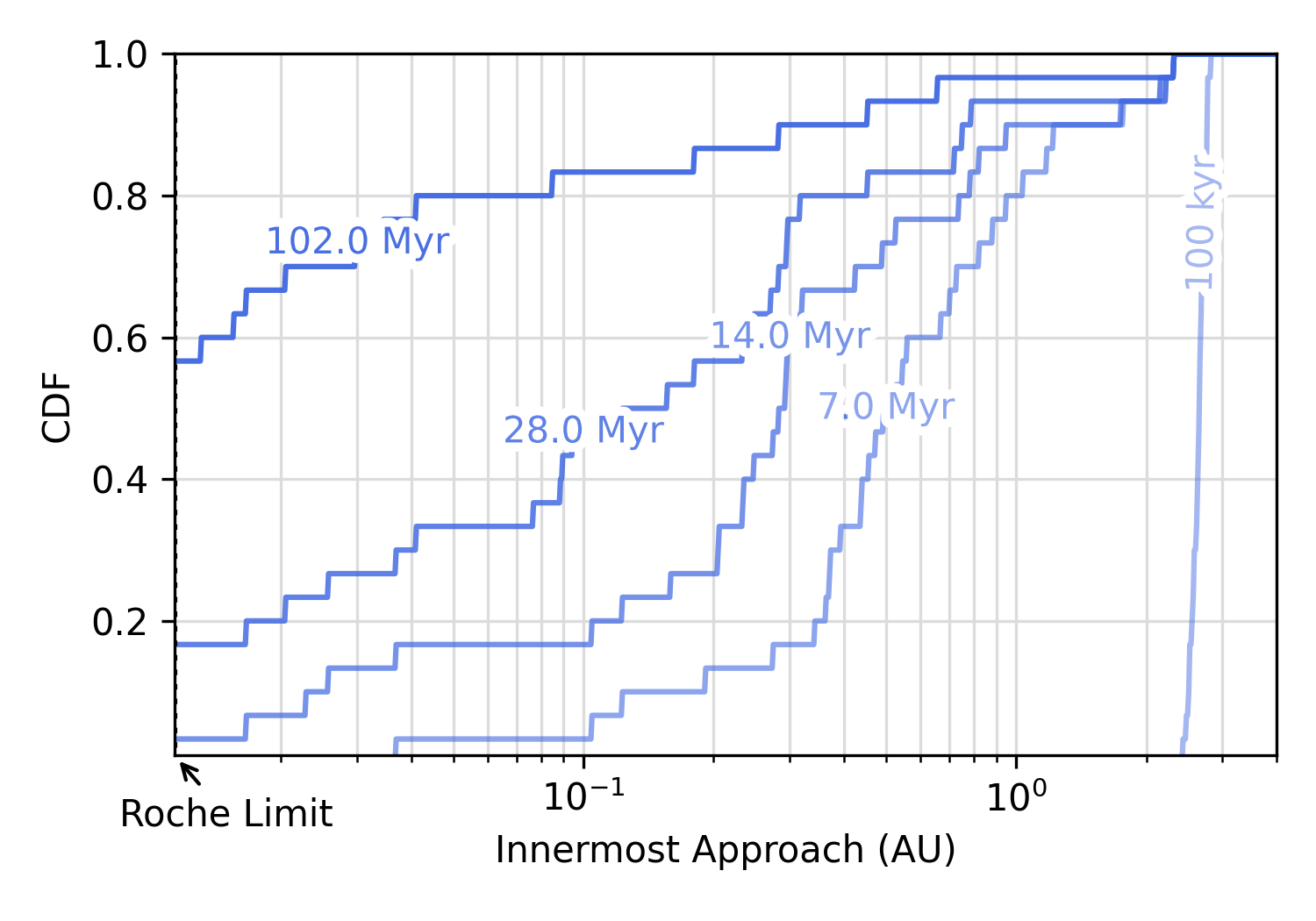}
\includegraphics[width=5.4cm]{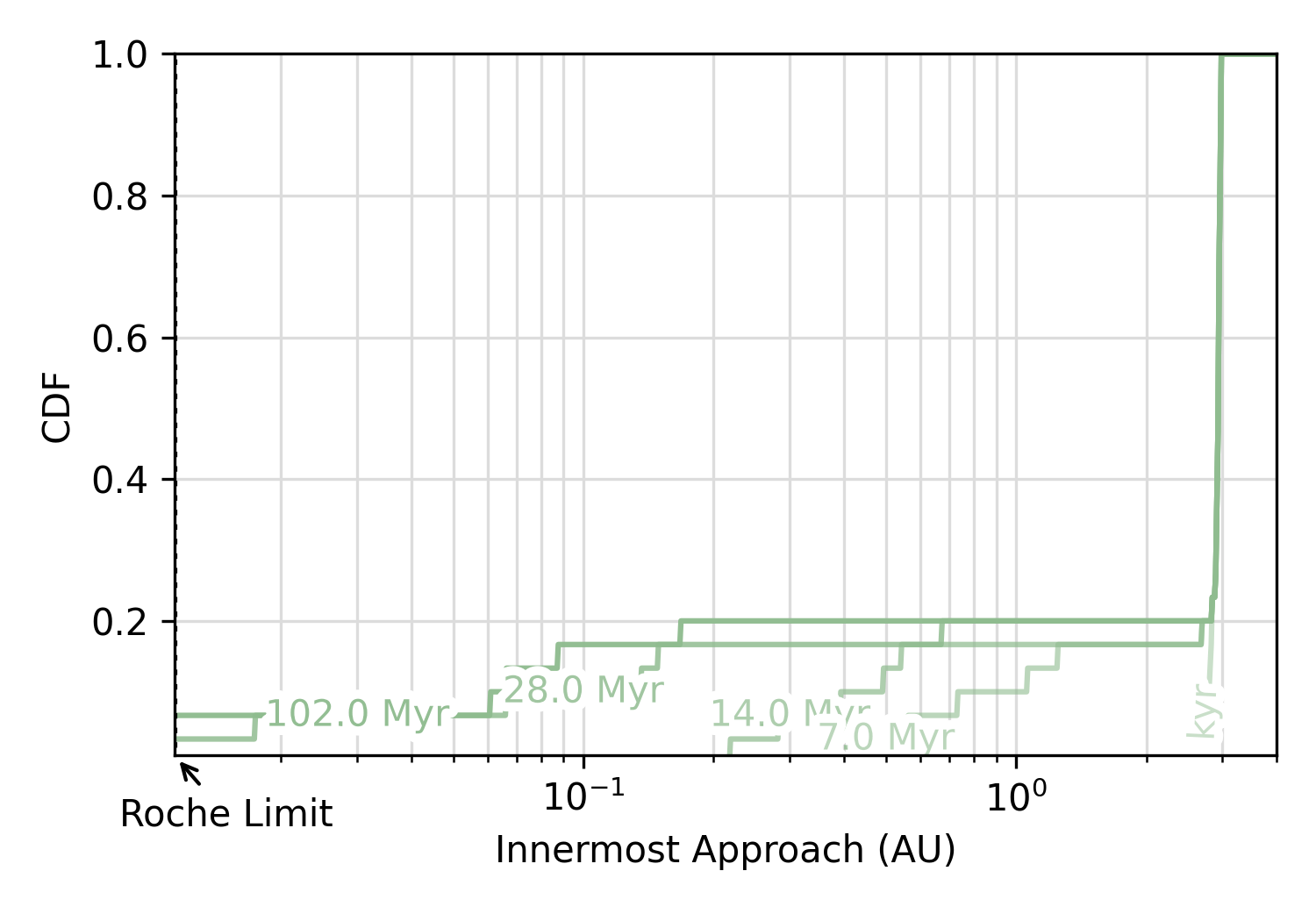}
\includegraphics[width=5.4cm]{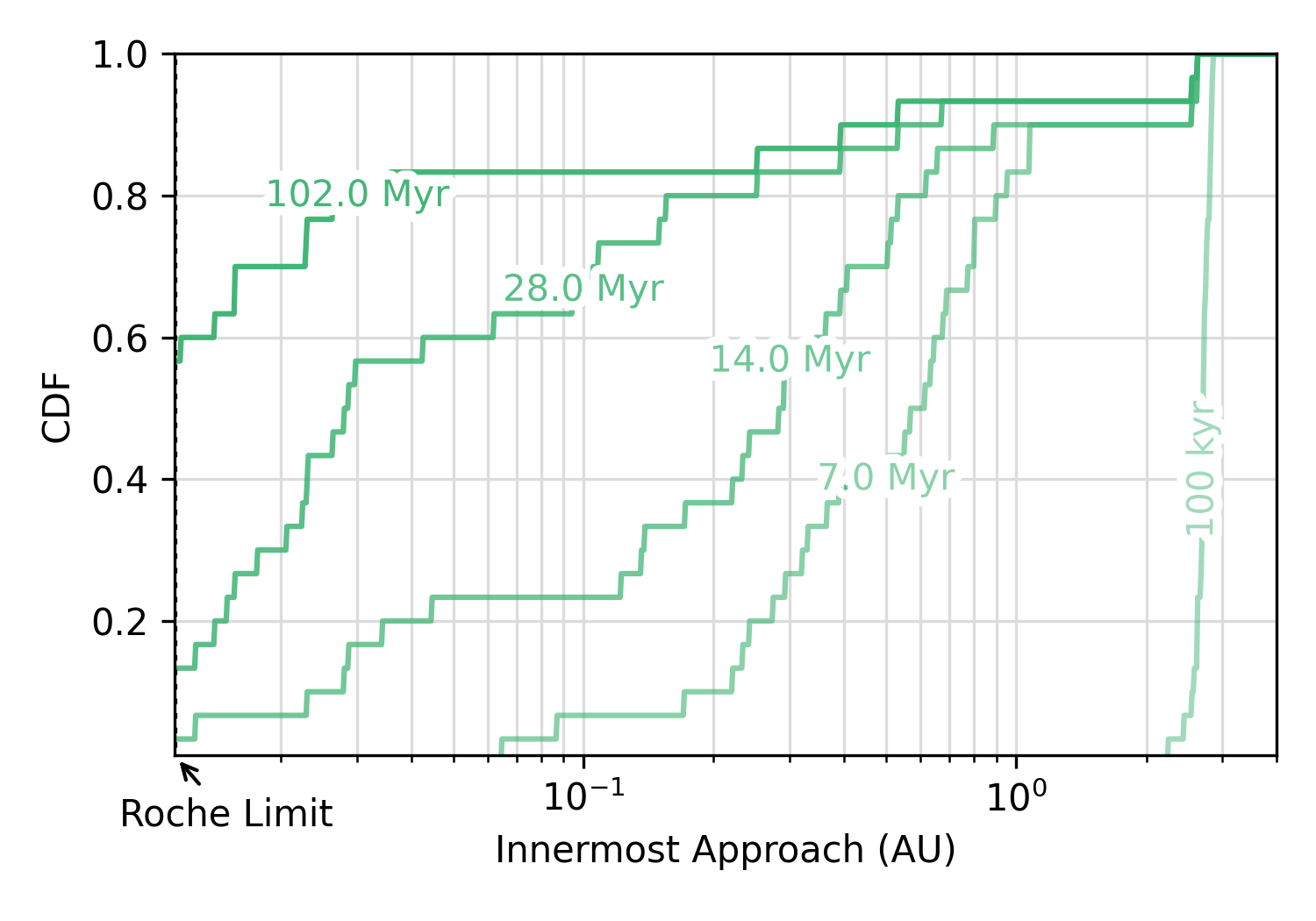}
\includegraphics[width=5.4cm]{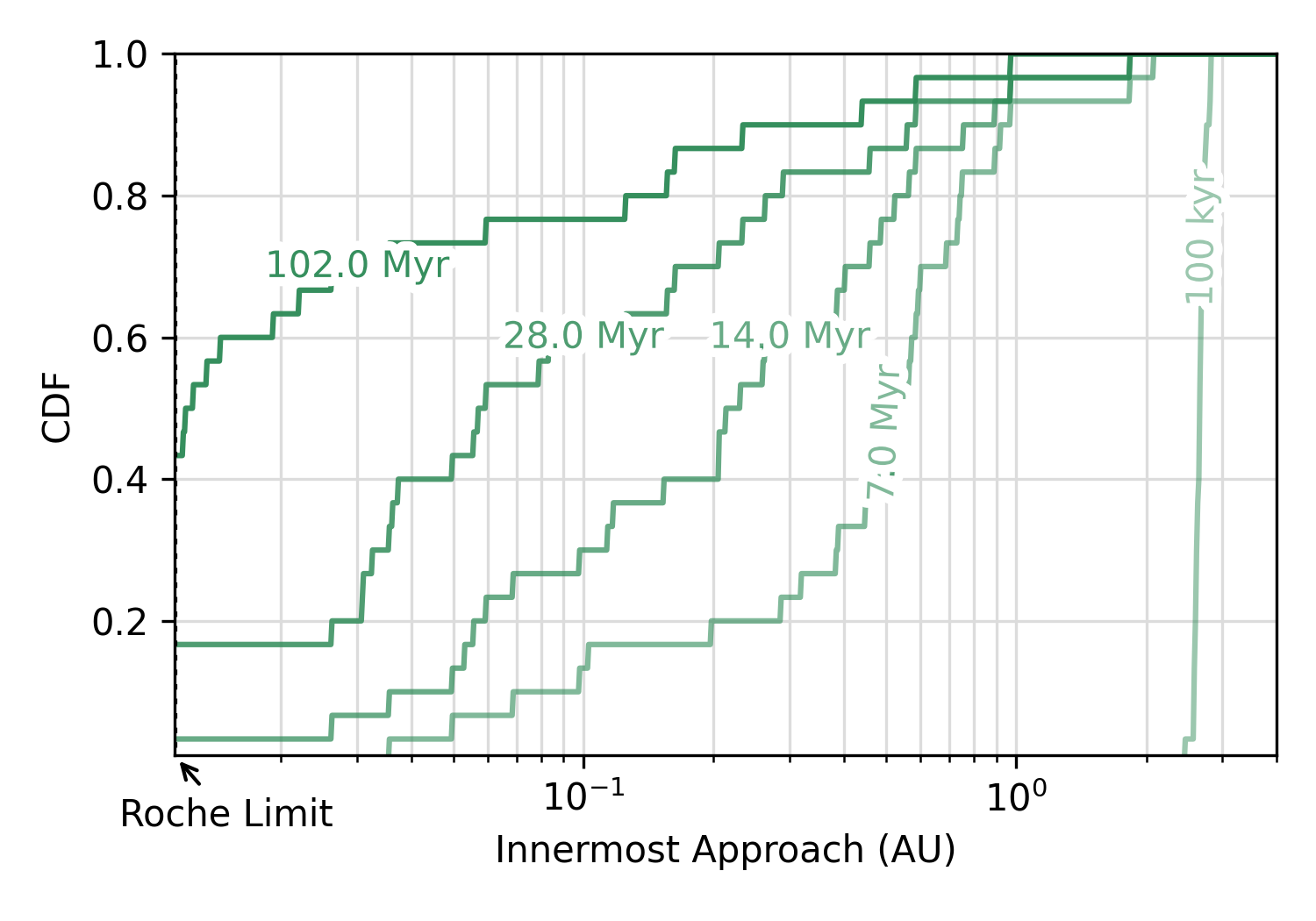}
\caption{As Figure~\ref{figure:close_approach}, but across all \textit{f5} suites of simulations. From left to right, top to bottom, these correspond to the $\textit{f5.3n1m}$, $\textit{f5.3n2m}$, $\textit{f5.3n4m}$, $\textit{f5.6n1m}$, $\textit{f5.6n2m}$, $\textit{f5.6n4m}$, $\textit{f5.12n1m}$, $\textit{f5.12n2m}$, and $\textit{f5.12n4m}$ simulations. In other words; total mass of planetesimals increases from 1\% to 2\% to 4\% from left to right, and the total number of planetesimals increases from 15 to 30 to 60 from top to bottom.}
\label{figure:planet_statistics}
\end{figure*}

Rates of inner system exploration subsequent to destabilization are similar across all \textit{f5} simulations\footnote{CDFs for the whole \textit{f5} series can be found in Figure~\ref{figure:planet_statistics}, located at the end of the paper.}. The distributions of estimated 0.1 AU crossing times and the proportion of systems containing a planet expected to drift below 0.1 AU for each of the \textit{f5} simulations are shown in Figure~\ref{figure:close_approach_times}. The time of approach within 0.1 AU for systems in our \textit{f5} series of simulations varies between approximately 10-40 Myr, with the median close approach time after destabilization being well-centered at 20.7 Myr. The proportion of total systems which achieve destabilization and this close approach is notably lower in the \textit{f5.12n1m} and \textit{f5.6n1m} suites; this is likely an effect of the high number of stable systems in those suites. The proportions of destabilized systems which achieve this close approach in those suites are $5/6 \sim 83\%$ and $17/22 \sim 77\%$---in accordance with the rest of the population.

On the other hand, suites of simulations with different planetary masses produce starkly different timescales for approach to the inner system. 
The \textit{mm(-3)f5}, or sub-Neptune series, and the \textit{mm(+3)f5}, or super-Neptune series, display median timescales for approach within 0.1 AU of 120 Myr and 4 Myr respectively. 
We therefore remark that the time from destabilization to first approach to the inner system is primarily a function of planetary mass and not planetesimal disk parameters.

\begin{figure*}[!tb]
\centering
\includegraphics[width=18cm]{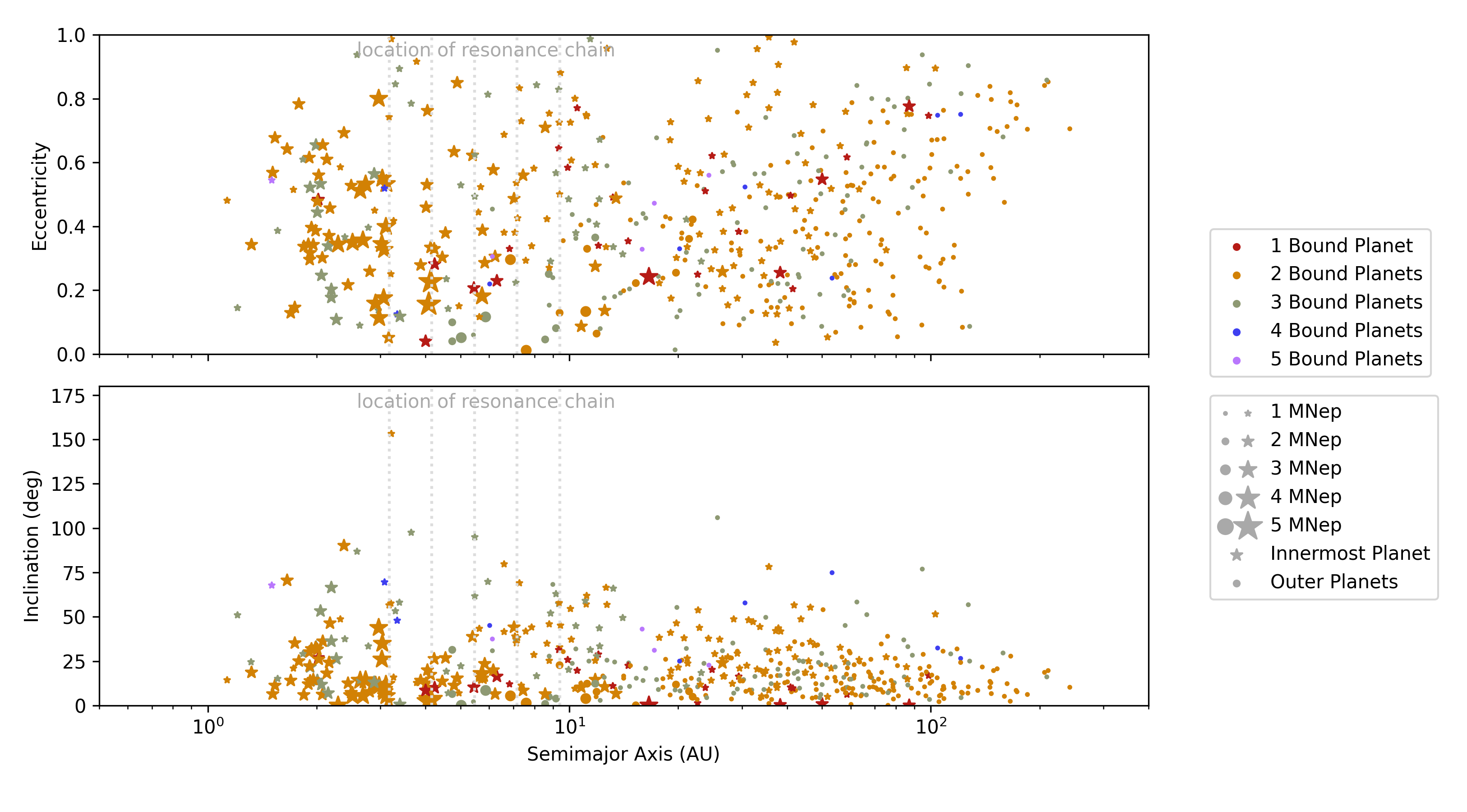}
\caption{Final eccentricities (upper panel) and inclinations (lower panel) as a function of semimajor axis for Neptunes in the \textit{f5} series that remain bound after 1.146 Gyr across all simulation suites. Colors indicate the system multiplicity, symbol size scales with planetary mass, and stars denote the innermost planets. Only destabilized systems are shown; the initial semimajor axes are indicated by vertical dotted lines.}
\label{figure:mass_seg_old}
\end{figure*}

\subsection{Mass Segregation}
\label{subsec:ms}

Figure~\ref{figure:mass_seg_old} shows the distribution of surviving Neptunes in semimajor axis-eccentricity space and semimajor axis-inclination space. 
The remaining bound Neptunes can be split into two populations based on their clustering: higher-mass Neptunes tend to cluster at lower semimajor axes while lower-mass Neptunes tend to cluster at higher semimajor axes, with the average eccentricity of the smaller Neptunes rising as their semimajor axis does. This produces a `V'-shaped clustering pattern in semimajor axis-eccentricity space characteristic of prior scattering experiments \citep{Chatterjee,beauge_nesvorny2012,Petrovich2014}, albeit here with significant scatter in semimajor axis, possibly due to the wider range in semimajor axis from which our planets are destabilized. 
Most Neptunes in our simulations have inclinations of less than 50 degrees, and no apparent difference emerges between the populations in inclination.

The high-mass population is gathered around the collisionless minimum energies for the system. For collisionless 1 $M_{Nep}$ Neptunes on their initial orbits, the minimum energy should lie at around $\sim 1$ au. Neptunes of a given mass tend to cluster around their minimum energies---this is most visible in the case of the 2 $M_{Nep}$ (at $\sim 2$ au) and 4 $M_{Nep}$ planets (at $\sim 4$ au) in this population. Clustering of massive planets at minimum energies is consistent with the results of \citet{HaddenWu2025}.

\begin{figure}[!tb]
\centering
\includegraphics[width=9cm]{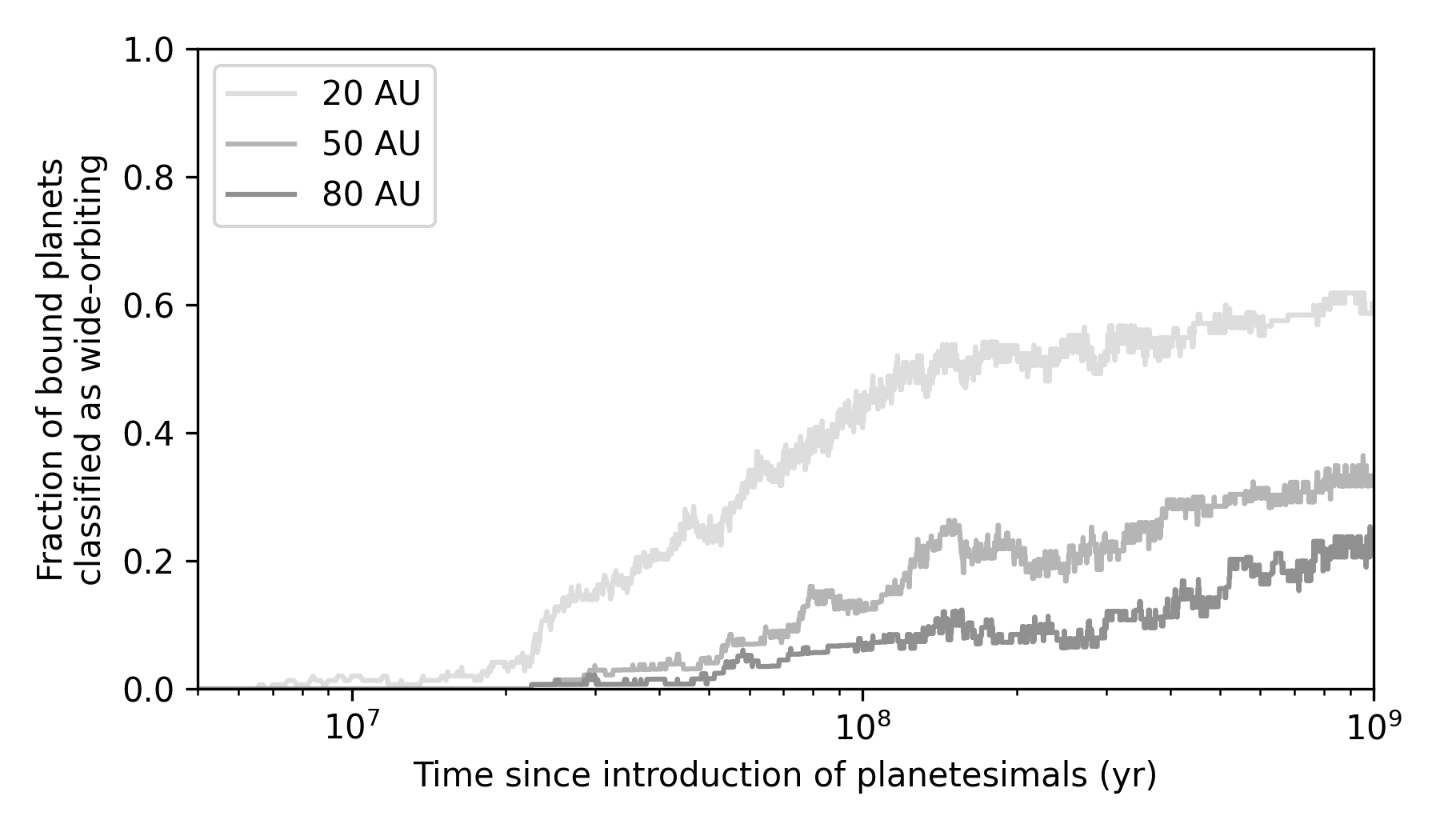}
\caption{Fraction of Neptunes classified as wide-orbiting above a given threshold in average separation $\langle r \rangle$ as a function of time for the fiducial \textit{f5.6n2m} simulations. Different thresholds are indicated by curves with varying opacity.}
\label{figure:wide_orbiting}
\end{figure}

Our simulations also produce a significant number of wide-orbiting Neptunes. The low-mass population is gathered around pericenters of 10 to 50 AU; correspondingly many of these Neptunes have orbit-averaged separation $\langle r \rangle$ above tens of AU, with
\begin{equation}
    \langle r \rangle = a \left(1+\tfrac{1}{2}e^2\right)
\end{equation}
(see e.g., \citealt{Tremaine2023}). Figure~\ref{figure:wide_orbiting} indicates the fraction of bound planets which can be classed as wide orbiting in our fiducial simulation. Even for definitions of wide-orbiting as distant as an orbit-averaged separation of 80 AU, near 25\% of bound planets in our fiducial simulation would be considered wide-orbiting.

We explain the distinct populations observed in Figure~\ref{figure:mass_seg_old} in the following way.
In a set of planets in which encounters repeatedly occur which are insufficient to eject or disrupt a planet, if there is a small probability of collision for a given such encounter over a sufficiently high number of such encounters collision will occur. 
Again from Safronov arguments (see \S\ref{subsec:cde}) we expect that these collisions occur predominantly (but not exclusively) among the inner planets of a set. 
When a now more massive planet interacts with one of the remaining, less massive planets, scattering outcomes change considerably as compared to an equal-mass or near equal-mass case. 
\citet{ford_rasio2008} find that mass is the primary determinant of ejection in a close encounter; mass ratios of 2:1 or above eject the lighter planet 99\% of the time, irrespective of initial semimajor axes, whereas mass ratios of 1:1 eject both planets at roughly similar rates.
As Neptunes in our simulations only gain substantial mass from collisions with other Neptunes, this has profound implications for the mass distribution of our evolved systems.
A `mass segregation' effect thus arises from the tendency for collisions to occur on average before scatterings, which thus produces an inner population of massive ($\geq2~\rm M_{Nep}$) collided Neptunes and an outer, scattered population of Neptunes of the original mass.

The prevalence of innermost planets remaining in Figure~\ref{figure:mass_seg_old} outwards of 20 AU and thus in the low-mass/scattered population is also of note. In these systems, all planets interior were lost due to a combination of collisions, disruptions, and ejections, ending with a disruption. In some cases, only one Neptune remains in the system as a wide-orbiting planet. We remark that in order for these planets to have been produced, some mechanism other than a close encounter likely must have driven the eccentricity growth to produce a disruption.

\subsection{Relaxed Systems}
\label{subsec:relax}

After 1.102 Gyr of evolution, most systems relax to 2-3 eccentric cold Neptunes. 
However, the high number of systems in the \textit{f5} simulations still undergoing substantive dynamical evolution in the prior 100 Myr after over 1 Gyr of evolution (96/270) indicates that some systems may not yet be stable.
Correspondingly, we examine 2-planet systems for stability as they constitute most systems in our \textit{f5} simulations by 1.102 Gyr.

\citet{Petrovich2015} provides stability criteria for two-planet systems of 
\begin{equation}
    r_{\rm ap} = \frac{a_{\rm out} (1-e_{\rm out})}{a_{\rm in} (1+e_{\rm in})} > Y
\end{equation}
for some $Y$ setting the criterion and semimajor axes, eccentricities, and planet-to-star mass ratios $a_i, e_i,\mu_i$ of the ith planet. 
A value of $1$ for Y, for example, sets the criterion such that there be no orbit crossing between planets irrespective of orientation. 

By SVM fitting to best produce $Y$ from simple terms, \citet{Petrovich2015} provide the criterion 
\begin{equation}
    Y_{\rm P15} = 1.15 + 2.4 (\textrm{max}(\mu_{\rm in},\mu_{\rm out}))^{1/3}(a_{\rm out}/a_{\rm in})^{1/2}
\end{equation}
When applied to our fiducial suite, we find that 10/24 two-planet simulations do not satisfy the stability criterion\footnote{This $Y$ works best for systems with mutual inclinations $\leq40$\textdegree, which most of our systems possess.}.
Of these 10, one pair is very nearly co-orbital (differing by less than one part in 100 in semimajor axis), three are non-co-orbital but regularly interacting, and six have sufficiently eccentric orbits that orbit crossing is possible in some orientations. 
Applying the $Y_{P15}$ criterion across all suites, we find that the rate of instability among 2-planet systems is similar across suites of simulation. Among relaxed systems, the median ratio $a_{out}/a_{in}$ is significant, at approximately 6.6 in the fiducial case, and ranges beteen $\sim4-12$ depending on the simulation suite chosen.
The prevalence of systems which are either still undergoing substantive evolution or which do not fit stability criteria indicates that about 40\% of systems are not yet fully relaxed. 
Stability criteria from \citet{EK95} and \citet{GMC13}, (with $Y_{EK95}$ and $Y_{GMC13}$ provided by \citet{Petrovich2015}) perform similarly.

\section{Discussion} \label{sec:discussion}
 
Our $N$-body experiments show that systems of cold Neptunes initially assembled into resonant chains can be destabilized by modest planetesimal disks containing only $\sim1$--$4\%$ of the total mass of the planets (of order an Earth mass). Cumulative scattering with planetesimals gradually breaks the resonances and increases the orbital spacing and eccentricities of the planets, ultimately triggering a global dynamical instability that reshapes the architecture of the system. The onset of the instability spans a wide range, from $\sim1$ Myr to Gyrs, reflecting a strong sensitivity to both the total mass of the planetesimal disk and its mass spectrum.

The main outcomes of the instability can be summarized as follows:

\begin{itemize}

\item {\it Sequence of dynamical outcomes.}
Following the instability, systems typically first undergo planet-planet collisions. These are followed on longer timescales by close approaches to the host star, leading to tidal disruptions and by ejections.
Remarkably, in most realizations at least one planet undergoes tidal disruption by the host star.

\item {\it Destruction of compact inner systems.}
If compact inner planetary chains are present, close-approaching Neptunes would interact with and likely destroy those chains on timescales of $\sim30$--$1000$ Myr. This delay is set both by the timescale for planetesimal-driven resonance breaking and by the gradual secular drift of the Neptune's pericenter toward the inner system.

\item {\it Final system architecture and wide-orbit populations.}
After instability, systems typically relax into mass-segregated configurations hosting two surviving planets, one of which often resides on a wide orbit ($\gtrsim 50$ au).
Because tidal disruptions remove planets from the inner system, some outcomes consist only of planets on wide orbits. 

\end{itemize}

In the following subsections we place these results in context by comparing with previous work, discussing the observational implications of this instability channel, and outlining the main caveats of our modeling.

\subsection{Breaking chains due to planetesimal disks}

Planetesimal-driven scattering as a means for planetary migration has been a well-studied feature of planetary dynamics for over forty years. 
Early applications involved investigations of Solar System giants: experiments embedding Uranus and Neptune in a planetesimal disk through scattering interactions drive Jupiter inwards and Saturn, Uranus, and Neptune outwards \citep{Fernandez1984,Hahn1999}. 
Outwards orbital migration of Neptune through resonant capture produces populations of eccentric planetesimals at resonances \citep{Malhotra93,Malhotra1995}; reproducing observed orbital parameters of resonant KBOs requires a planetesimal disk mass of roughly $\sim 50 M_\Earth$ to produce the requisite planetary migration \citep{Hahn1999}.
Later models, including that of \citet{Nice2}, reproduce the positions, eccentricities, and inclinations of the Solar System's giant planets from planetary migration induced by a $\sim 30$-$40 M_\Earth$ planetesimal disk resulting in a 2:1 MMR crossing between Saturn and Jupiter\footnote{Crossing such a resonance from below results in jumps in eccentricity and separation~\citep{henrard1983,Wu_pingpong2024}.}.

Planetesimal-driven dynamics have also used to explain the properties of exoplanet populations. 
The sample of Kepler planets exhibits asymmetry in period ratios between planets near first-order MMRs, with planets tending to `pile-up' just past the MMR leaving a dearth of planets beneath it~\citep{Lissauer2011}.
\citet{LithwickWu2012} use ``resonant repulsion", a mechanism by which pairs of planets near MMRs increase in separation in the presence of eccentricity damping (such as in the case of a planetesimal disk; see e.g., \citet{Murray2002}) to reproduce the Kepler asymmetry about MMRs. 
Scattering of planetesimals from disk masses as small as $\sim 2 M_\Earth$ to $\sim 50 M_\Earth$ has been invoked as a means to break resonance and increase period ratios between Neptunes and lower masses planets and reproduce this asymmetry~\citep{Chatterjee2015,Ghosh2023_repulsion}.

\citet{Wu_pingpong2024} established the mechanism behind this increase: for an isolated pair of planets and a single planetesimal scatterer, the scatterer preferentially absorbs angular momentum over energy from the pair, increasing their separation in semimajor axis--a process they denote ``ping-pong repulsion". 
In the case of many scatterers, the Kepler asymmetry can be explained by as little as 1\% of the mass of a pair of planets in planetesimals. 
Expanding from pairs of planets to resonance chains, \citet{HaddenWu2026} and \citet{Choksi2026} study breaking inner resonance chains of super-Earths with planetesimals in-situ. In their simulations, destabilization requires planetesimal disk masses between $\sim 1$-$5 M_\Earth$ in the inner system.  Our work is complementary as it expands the range of planetesimal disks and explores cold Neptunes.

In our suites of simulations, we use comparatively low-mass disks and high-mass individual perturbers. Our disks range between $\sim 1-3.5 M_\Earth$ (with one $\sim 10 M_\Earth$ disk in the \textit{mm(+3)f5} series). We use substantially larger individual planetesimals than \citet{HaddenWu2026}, whose planetesimals are at most $\sim12\%$ the mass of those in our fiducial case. Our fiducial planetesimals are more comparable to the largest used in the scattering experiments of~\cite{Chatterjee2015} ($\sim45\%$ the mass of our fiducial planetesimals). The least massive planetesimals in each case are $\sim 4\%$ and $\sim1.5\%$ the mass of those in our fiducial case, respectively. On the other hand, our fiducial planetesimals are substantially smaller those of \citet{Choksi2026} ($\sim 300\%$ the mass of our fiducial planetesimals).

A prerequisite for these mechanisms is the existence of the planetesimal disks. While others invoke Mercuries in the inner regions ($\sim 0.1 $ au), we invoke similar and smaller masses but at much greater orbital separations ($\sim 10$ au). The presence of planetesimal disks akin to those we invoke can be inferred from debris disk observations; observed debris disks are typically centered at tens to hundreds of AU~\citep{Holland2017,Matra2025}.
Constraining debris disk sizes on the low end by mass loss rates and on the high end by the mass of solids in typical protoplanetary disks, \citet{Krivov2021} surmise that in general debris disk masses should be constrained to between $10 M_\Earth$ and $100$s of $M_\Earth$; however, if an idealized planetesimal size distribution is used, observations systematically predict debris disk masses roughly an order of magnitude higher ($1000$s of $M_\Earth$). The authors note that if planetesimals $>1\, 
\rm{km}$ in radius in outer systems are rarer than predicted, observationally implied masses would align with their limits. This may have implications for the quantity of massive planetesimals in our disks.

\subsection{The timescale to break the chains of inner transiting planets}
\label{subsec:timescales}

Observational evidence has begun to accrue that indicates planetary systems form in resonance chains which dissolve on timescales of the order of 100s of Myr. 
\citet{Dai2024} find that transiting multiplanet systems less than 100 Myr in age typically contain at least one resonance or near-resonance between planets.,  
and that the fraction of systems in resonant or near-resonant configurations drops sharply with age.
Additional evidence for the dissolution of resonant configurations comes from measurements of transit timing variations (TTVs), whose amplitudes and thus detectability increase dramatically near resonance~\citep{Agol2005}. 
Among transiting multiplanet systems less than 800 Myr in age\footnote{The young-system sample of \citet{Lopez2026_TTV} leaves out some known young systems with TTVs, which may increase the proportion of systems with them.}, the proportion of systems with TTVs is elevated by roughly a factor of four~\citep{Mazeh2013,Lopez2026_TTV}. 

The frequency of resonance chains in young systems matches some predictions of \citet{Izidoro2017}'s ``breaking the chains" model, in which many planets are formed in the disk, convergently migrate into close resonances, and upon disk dispersal this `overpacked' resonance chain dissolves due to collisions between the planets.
In the breaking the chains model, most systems have all their collisions occur within 5-20 Myr~\citep{Izidoro2017,Izidoro2021_breaking}. 
The timescale of the breaking the chains model implies that the timescale for dissolution of resonance chains is too long to be explained primarily by collisions from overpacking after disk dispersal.

A variety of mechanisms for breaking inner resonance chains due to the effects of outer, unseen bodies have recently been proposed. 
\citet{Goldberg2025} and \citet{Ogihara2026} evolve many small planetary embryos in disks of $10$-$40M_\Earth$ from near 1 AU, producing stable inner chains, with residual populations of small planets ($\lesssim 1 M_\Earth$), in an outer chain at around 1 au, which if destabilized can trigger a rapid destabilization of the inner chain in 10s to 100s of Myr.
Dynamical interactions in the outer chain can produce secular perturbations exciting the inner chain on similar timescales. 
Instead of higher-mass perturbers, \citet{intruder2026} proposed the breaking of inner chains from numerous or repeated flybys of low-mass perturbing planetesimals, projecting timescales of 100s of Myr for $m_p \sqrt{N_{flyby}} \sim 50$-$100 M_\Earth$. 
In contrast, \citet{HaddenWu2026}'s in-situ planetesimal breaking finds timescales spanning from $<1~\rm Myr$ up to over 100 Myr, with most timescales for destabilization being around 10 Myr, albeit for a limited sample of simulations. 

We propose that outer Neptunes (or sub-Neptunes) can drive instabilities in the inner system on the timescales required. 
In our model, following an instability in a resonance chain of the outer system, planets in that chain can reach inwards of $\sim0.1-0.2$ AU, thereby destabilizing stable resonance chains in the inner system. 
However, we argue that the timescale for instabilities of resonance chains of Neptunes is too sensitive to planetesimal disk parameters to provide a universal explanation for the breaking of chains in a 100 Myr timescale; as mentioned in \S\ref{subsec:destab_t}, we find that distributions of destabilization time can span up to two orders of magnitude. 
Instead, we find it more plausible that timescales of gradual, chaotic drift to 0.1 au are robust to changes in disk parameters. 
Such timescales depend primarily on the masses of the interacting planets driving the inwards drift. 
If planetesimal disks are sufficiently massive ($\gtrsim 4 M_\Earth$, so destabilization timescales of the outer system are short) and planets in the outer system are super-Earths or larger, a 100 Myr timescale for breaking the chains of inner transiting planets would be consistently observed.

\begin{figure}[!t]
    \centering
    \includegraphics[width=8cm]{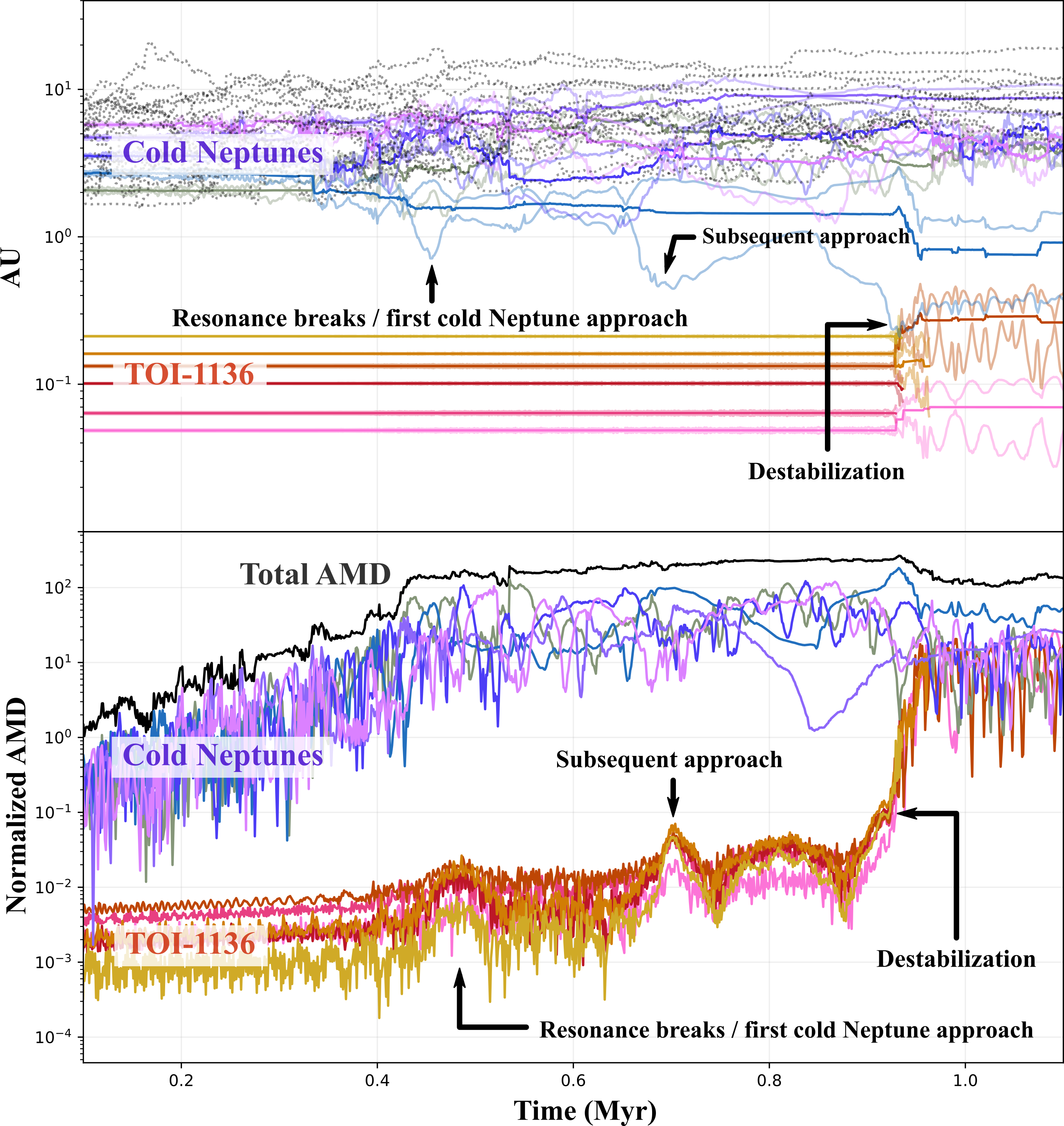}
\caption{Proof-of-concept evolution of an inner resonant chain of super-Earths coupled to a destabilizing outer resonant chain of cold Neptunes. The super-Earths are initialized with the masses and resonances of the TOI-1136 system, except for the 7:5 resonance (planets e-f) that is replaced by a 4:3 resonance to enhance its stability. Colored lines trace the orbital evolution of each planet, with warm colors denoting super-Earths and cool colors denoting cold Neptunes. The simulation is selected such that the outer chain destabilizes within $\sim 200$ kyr and the inner chain within $\sim 1$ Myr. 
{\bf Top:} Evolution of semimajor axis, pericenter, and apocenter for each planet; dashed black lines indicate the time-averaged positions of planetesimals. 
{\bf Bottom:} Evolution of the angular momentum deficit (AMD), normalized to its value at the time planetesimals are introduced. Key events in the evolution of the inner resonant chain are labeled in both panels.}
    \label{fig:toi1136}
\end{figure}

In our \textit{f5} simulations we conducted integrations of only the orbits of outer cold Neptunes. 
In order to evaluate the ability of this mechanism to destabilize inner chains, we have also conducted a set of preliminary simulations in which an outer resonance chain of cold Neptunes is evolved alongside an inner resonance chain of super-Earths. 
A characteristic example is shown in Figure~\ref{fig:toi1136}. 
The top panel tracks the orbital evolution of the system for 1 Myr after disk dispersal and the bottom panel tracks the evolution of the AMD. 
For these preliminary simulations, the inner resonance chain approximates the orbital distances and period ratios of TOI-1136 as reported in \citet{dai2023}. 

The parameters of the system displayed in Figure~\ref{fig:toi1136} are chosen so as to decrease destabilization time and time between destabilization and close approach---an induced `fast instability' as compared to our \textit{f5} series of simulations. As the innermost cold Neptune's eccentricity is excited and its perihelion approaches the inner chain, the AMD of the inner chain temporarily rises before relaxing to only a slightly more elevated level. Once a cold Neptune passes within about 0.2 AU, becomes orbit-crossing or very nearly so, the inner six-planet resonant chain rapidly destabilizes, collapsing to two eccentric close-orbiting planets within a few tens of thousands of years. 

Moreover, the close approach of a Neptune to within 0.1 to 0.2 AU may be excessive. Formal breaking of the resonance\footnote{The distinction drawn in \citet{Dai2024} and others is between the resonant systems whose resonant angles librate and the near-resonant systems whose resonant angles circulate; see also the discussion in the last paragraph of \S2 of \citet{Goldberg2023}.} labeled in Figure~\ref{fig:toi1136} occurs in our preliminary simulations after a Neptune approaches within approximately 1 AU, well before the apparent (based on period ratios) breaking of the resonance chain and destabilization at a 0.1-0.2 AU approach. A long-term stable resonant system, which subsequent to a distant (1 AU) approach becomes near-resonant, may not exhibit the same stability in its near-resonant state. Further work is required to understand this paradigm of system.

\subsection{Cold Neptunes: enhanced tidal disruptions by secular chaos}
\label{subsec:cn_disrupt}

A notable outcome of our simulations is the prevalence of systems in which a Neptune approaches the host star to within $\sim2\,R_\odot$. In our fiducial set, we find 35 such events across 30 systems (see Table~\ref{table:summary_final}). At such distances, we treat the planet as tidally disrupted; if tidal dissipation were included, some of these passages could instead lead to tidal capture, an effect we do not model here.

We can compare these results with previous planet--planet scattering experiments that have primarily focused on higher-mass, Jovian-like planets \citep{Chatterjee,JT2008}. Simulations that include tidal capture find that systems of three Jupiters produce tidal captures in $\sim10$--$30\%$ of cases \citep{nagasawa2008,nagasawa2011,beauge_nesvorny2012,CP_review2024}, while four-Jupiter systems yield capture fractions of $\sim20\%$ \citep{beauge_nesvorny2012}. These rates are a factor of $\sim3$--$10$ lower than those inferred from our Neptune-mass simulations.

We suggest that the higher incidence of tidal disruptions (and correspondingly fewer ejections) at lower planetary masses is a consequence of eccentricity excitation driven by {\it secular chaos}. As illustrated in Figure~\ref{figure:destab}, during the instability phase the innermost planet undergoes extended episodes of eccentricity growth at nearly constant orbital energy (i.e., constant semimajor axis), while the outer planets continue to exchange energy and exert torques on the inner orbit. This process can persist for long timescales as long as the outer planets remain bound.

All else being equal, the characteristic timescale for secular torquing scales inversely with planetary mass ($\tau_{\rm secular} \propto M_p^{-1}$), whereas the timescale for energy diffusion leading to ejection scales more steeply ($\tau_{\rm ejection} \propto M_p^{-2}$) for test particles scattered by a massive planet \citep{Tremaine1993}. Numerical experiments suggesting a somewhat shallower dependence ($\tau_{\rm ejection} \propto M_p^{-1.64}$) \citep{HaddenWu2025}. A useful proxy for the efficiency of secular chaos is therefore the ratio of these timescales, which measures the available time for the system to explore phase space and reach extreme eccentricities:
\begin{equation}
\frac{\tau_{\rm secular}}{\tau_{\rm ejection}} \propto M_p^{0.64\text{--}1}.
\end{equation}

This scaling implies that lower-mass systems have relatively more time to reach extreme eccentricities before ejection, enhancing the likelihood of tidal disruption or capture. Accordingly, the ratio of tidal disruptions to ejections is expected to decrease with increasing planetary mass. In our fiducial Neptune-mass simulations we find $D/E = 35/32 \simeq 1.1$. Repeating the same experiment with Jupiter-mass planets yields $D/E = 13/41 \simeq 0.32$. This trend is consistent with the expectation that higher-mass systems are more efficient at ejecting planets before they can reach extreme eccentricities. A more detailed exploration of this dependence is left for future work.

\subsubsection{Hot Neptunes by high-eccentricity migration}
An important implication of our results is the formation of hot Neptunes through high-eccentricity migration. This pathway is analogous to the secular chaos scenario proposed for hot Jupiters \citep{Wu_secularchaosHJ2011}, in which long-term interactions among multiple planets drive extreme eccentricities followed by tidal circularization. Observational evidence supports this picture: an emerging population of hot Neptunes with orbital periods of $\sim3$--$6$ days exhibits large stellar obliquities \citep{Dai2017} and, in some cases, residual eccentricities \citep{Hixenbaugh2023,Dugan2025}, both indicative of a dynamically excited origin \citep{Dawson_HT_review,Bourrier2023_hot_Neptunes}.

The excitation of such extreme eccentricities is typically attributed to massive outer companions, such as cold Jupiters or stellar binaries. However, long-term radial-velocity monitoring rules out the presence of cold Jupiters in most hot Neptune systems \citep{Espinoza-Retamal2024}. Neptune-mass companions at wide separations would often remain undetected. Our simulations naturally produce such architectures: systems undergoing dynamical instabilities frequently retain one or more Neptune-mass planets on wide orbits, which can continue to secularly excite the inner planet. In our fiducial simulations, the outer companion to a tidally disrupted planet typically resides at $\sim 20$~au, placing it beyond the reach of current detection limits.

Given the high occurrence rate of Neptune-mass planets and the relative scarcity of cold Jupiters, Neptune--Neptune interactions provide a natural and likely dominant pathway for driving the eccentricity excitation required for high-eccentricity migration of Neptunes. In this picture, hot Neptunes are a natural outcome of dynamical instabilities in multi-Neptune systems, with the required perturbers being wide-orbit Neptunes that are typically undetectable with current techniques. A more complete assessment of this channel will require coupling the dynamical evolution explored here with detailed tidal models, including temperature-dependent rheologies, which we defer to future work.

\subsection{Relaxed systems and microlensing demographics}

Dynamical instabilities among Neptune-mass planets naturally lead to mass-segregated configurations in which the most massive planets remain on relatively tight orbits while lower-mass planets are scattered to wider separations (see Figure \ref{figure:mass_seg_old}). 

One recent example is the candidate three-planet system HD~208487, which hosts a Saturn-mass planet at $\sim0.5$~au accompanied by two outer Neptune-mass companions at $\sim1.9$ and $\sim2.5$~au \citep{Rubenstein2025}. The authors show that such an architecture may arise from the instability of an initially long resonant chain of six planets locked in successive 3:2 resonances, each with masses of $\sim2$--$3\,M_{\rm Nep}$. 

This outcome closely resembles the relaxed configurations produced in our simulations, in which a small number of surviving planets remain bound while lower-mass planets are scattered to progressively wider orbits. Such systems may therefore represent the observational endpoint of dynamical relaxation in multi-Neptune systems. In particular, the presence of surviving wide-orbit Neptune-mass planets is a natural consequence of this process.

These wide-orbit planets are especially relevant for microlensing surveys, which are most sensitive to planets at separations of a few astronomical units or larger \citep{Gaudi_review}. Dynamical instabilities among cold Neptunes may therefore contribute to the population of wide-orbit planets inferred from microlensing and provide a natural link between compact resonant chains and the demographics of planets at large orbital separations.

A key prediction of our model is the relative abundances of cold planets, wide-orbit planets, and free-floating planets. Similar dynamical studies have explored this question, mostly focused on the production of free-floating planets (see, e.g., \citealt{Veras2012,Zhai2025,Bhaskar2025,Huang2026}). Our work complements these studies by considering a self-consistent framework to drive instabilities and by considering both planetary collisions and tidal disruptions. A subtle effect from the large incidence of tidal disruptions is a novel population of systems containing only wide-orbit planets, not captured in previous works.

Broadly, our results suggest an approximate equipartition of outcomes, with roughly one massive snow-line planet per ejection and per scattered wide-orbit planet (see Figure~\ref{fig:cde}). But this depends on what we define as wide-orbit. As shown in Figure~\ref{figure:wide_orbiting}, the fraction of wide-orbit planets decreases from $\sim 60\%$ for average separations of $\langle r \rangle > 20$ au to $\sim 20\%$ for $\langle r \rangle > 80$ au. Further modeling is required to quantify the extent to which wide-orbit planets are detected or instead classified as free-floating (see, e.g., \citealt{Han2005}). In particular, we aim to assess detectability for the Nancy Grace Roman Space Telescope microlensing survey \citep{Penny2019}.

\subsection{Missing physics and uncertainties}

In these initial suites of simulations, we have left out certain physical effects. We include neither prescriptions for tidal dissipation nor relativity-related precession effects. Our treatment of disk migration is intentionally kept as simple as possible. In addition, there are several aspects of our parameter space in which we have adopted simplifications that we intend to explore further in future work.

\paragraph{Resonance chain assembly} Here, we have chosen a fixed disk with uniform damping parameters ($\tau_a = 4\rm{~Myr}$, $K=100$) in order to produce deep 3:2 resonances. A more physical treatment of damping timescales of orbital elements for a range of disk parameters may produce different resonances among outer giants. 

From limited preliminary simulations, the 2:1 resonance seems too well-separated to consistently generate instability in the outer system for Neptune-mass planets under our considered parameters. On the other hand, higher resonances such as the 4:3 are more fragile; these preliminary simulations suggest five-planet resonance chains in the 4:3 destabilize much more quickly. Further investigation into resonance chain assembly in the outer system is required.

\paragraph{Peas-in-a-pod planets} In our simulations we have chosen to instantiate systems with equal-mass, peas-in-a-pod Neptune-like planets.  However, even small differences in mass ratios can have significant effects on the outcomes of close encounters~\citep{ford_rasio2008}. Correspondingly, a greater diversity of initial masses among the planets initially assembled in the chain ought to be considered. We also remark that our simulations have not considered modulations to the initial separations of the Neptune or sub-Neptune chains from the star which could have impacts for both destabilization and secular timescales.

\paragraph{Planetesimal disk} We assume planetesimal disks of uniform, massive planetesimals. This simplification omits ranges of mass, breadths, and locations of possible planetesimal disks as well as the size distributions of planetesimals in those disks, all of which may have a significant impact on system evolution given the sensitive dependence of destabilization time on planetesimal disk parameters. We intend to explore a larger range of planetesimal disk parameters using quasi-particles in future simulations.

Additionally, we implanted planetesimals into the system after the removal of the disk phase. In a more physically involved model, planetesimals could be instantiated with planets and evolved in the disk, subject to the effects of dynamical friction.

\section{Conclusions}
\label{sec:conclusion}

We present a model in which cold ($a \sim 3$--$10$ au) Neptunes (and sub-Neptunes) undergo a two-stage evolution: gas-disk migration assembles them into long resonant chains, which are subsequently disrupted by interactions with remnant planetesimal disks. This framework naturally produces a diverse range of outcomes that connect the observed demographics of short-period and wide-orbit planets.

\paragraph{Breaking resonance chains of close-in planets}
While planetesimal disks can break resonant chains and trigger global instabilities, we show that the associated timescales are extremely sensitive to disk mass and size distribution: factor-of-two variations produce order-of-magnitude changes. Long delay times from planetesimal scattering alone are therefore intrinsically fine-tuned to produce a sharp transition to instability (e.g., on $\sim 100$ Myr timescales as suggested by observations; \citealt{Dai2024}).

We instead identify a more robust pathway. Instabilities among cold $\sim 5\,M_\oplus$ planets naturally break inner resonant chains on a characteristic timescale of $\sim 100$ Myr. Secular chaos, driven by outer orbit-crossing evolution that ultimately leads to ejections, gradually drives the innermost planet to small pericenters. A robust outcome is the delivery of a planet to $\sim 0.1$--$0.2~\mathrm{au}$, where compact resonant systems are commonly observed.

Thus, if cold sub-Neptune systems become unstable within $\sim 100$ Myr (e.g., via a sufficiently massive planetesimal disk), inner chains are systematically disrupted on a predictable timescale that scales approximately linearly with planetary mass, owing to the dependence on secular chaos.

\paragraph{Production of high-obliquity hot Neptunes}
The fraction of unstable systems where a Neptune driven to pericenters of a few $R_\odot$ is remarkably high (at least $\sim 60\%$). Thus, in the absence of inner planets capable of quenching the gradual pericenter drift, high-eccentricity migration may efficiently produce hot Neptunes. This mechanism naturally explains the observed population of high-obliquity, apparently isolated hot Neptunes that lack detected outer Jovian companions. Instead, their migration may have been driven by outer Neptunes that remain below current detection limits.

\paragraph{Cold planet demographics and microlensing surveys}
Post-instability evolution, in which planet--planet collisions, tidal disruptions, and ejections occur at comparable rates, drives systems toward mass-segregated architectures. Lighter planets are preferentially ejected or scattered onto more loosely bound orbits at $\sim 50$--$100$ au, while more massive planets remain at a few au. 
Such architectures provide a direct observational test of cold-Neptune instabilities: microlensing surveys can probe the predicted branching ratios between cold Neptunes, very cold Neptunes, and a population of frigid free-floating planets.

\medskip

In summary, our results indicate that cold Neptunes (and sub-Neptunes) are not passive remnants of planet formation; rather, they play a central role in shaping planetary system architectures and the global demographics of exoplanets.

\section*{Acknowledgements}

We thank Antranik Sefilian for the many insightful discussions and helping to shape this project in its early stages. We thank also Eugene Chiang for helpful discussion and his suggestion regarding separating timescales for close approaches and destabilizations (without which we might not have had Fig.~\ref{figure:close_approach_times} nor the conclusions contained therein). We would further like to thank Chris O'Connor, Kaitlin Kratter, Eric Ford, Michael Poon, and Songhu Wang for their valuable discussions and insights. 

This research was supported in part by Lilly Endowment Inc., through its support for the Indiana University Pervasive Technology Institute.

\software{\texttt{astropy} \citep{astropy1,astropy2,astropy5}, \texttt{colorcet} \citep{colorcetKovesi2015,colorcetPy}, \texttt{labellines} \citep{labellines}, \texttt{matplotlib} \citep{matplotlib}, \texttt{numpy} \citep{numpy}, \texttt{REBOUND} \citep{REBOUND}, \texttt{REBOUNDx} \citep{REBOUNDx}}.

\bibliography{biblio,biblio_CP}

\end{document}